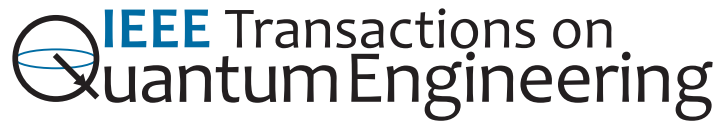

# Exploration of Design Alternatives for Reducing Idle Time in Shor's Algorithm: A Study on Monolithic and Distributed Quantum Systems

**MORITZ SCHMIDT[1], ABHOY KOLE[2], LEON WICHETTE[3], ROLF DRECHSLER[2,4] (Fellow, IEEE), FRANK KIRCHNER[1,3], AND ELIE MOUNZER[3]**

[1]Robotics Research Group, University of Bremen, 28359 Bremen, Germany
[2]German Research Center for Artificial Intelligence Cyber-Physical Systems, 28359 Bremen, Germany
[3]German Research Center for Artificial Intelligence Robotics Innovation Center, 28359 Bremen, Germany
[4]Group of Computer Architecture, University of Bremen, 28359 Bremen, Germany

Corresponding author: Moritz Schmidt (e-mail: moritz.schmidt@uni-bremen.de).

This work was supported by the German Ministry for Economic Affairs and Climate Action (BMWK) and the German Aerospace Center (DLR) through the project QuMAL-KI under Grant 50RA2208A (German Research Center for Artificial Intelligence) and Grant 50RA2208B (University of Bremen).

**ABSTRACT** Shor's algorithm is one of the most prominent quantum algorithms, yet finding efficient implementations remains an active research challenge. While many approaches focus on low-level modular arithmetic optimizations, a broader perspective can provide additional opportunities for improvement. By adopting a midlevel abstraction, we analyze the algorithm as a sequence of computational tasks, enabling systematic identification of idle time and optimization of execution flow. Building on this perspective, we first introduce an alternating design approach to minimize idle time while preserving qubit efficiency in Shor's algorithm. By strategically reordering tasks for simultaneous execution, we achieve a substantial reduction in overall execution time. Extending this approach to distributed implementations, we demonstrate how task rearrangement enhances execution efficiency in the presence of multiple distribution channels. Furthermore, to effectively evaluate the impact of design choices, we employ static timing analysis—a technique from classical circuit design—to analyze circuit delays while accounting for hardware-specific execution characteristics, such as measurement and reset delays in monolithic architectures and ebit generation time in distributed settings. Finally, we validate our approach by integrating modular exponentiation circuits from QRISP and constructing circuits for factoring numbers up to 64 bits. Through an extensive study across neutral atom, superconducting, and ion trap quantum computing platforms, we analyze circuit delays, highlighting tradeoffs between qubit efficiency and execution time. Our findings provide a structured framework for optimizing compiled quantum circuits for Shor's algorithm, tailored to specific hardware constraints.

## I. INTRODUCTION

Shor's algorithm [1] offers an exponential speedup for integer factorization, presenting a substantial threat to RSA-based cryptographic security [2], [3], [4], [5]. Although large-scale quantum computers capable of breaking RSA are still being developed, improving the efficiency of Shor's algorithm remains vital for advancing practical quantum computing and shaping postquantum cryptography research.

Most previous research on implementing Shor's algorithm has focused on optimizing low-level modular arithmetic circuits, particularly modular addition and multiplication [5], [6], [7], [8], [9], [10], [11], [12], [13], [14], [15]. However, these efforts primarily refine individual operations without fully exploiting the algorithm's overall structure to reduce execution bottlenecks. Adopting a midlevel abstraction enables us to break the algorithm into structured computational tasks rather than solely optimizing gate-level arithmetic. This task-based perspective facilitates a parallelization strategy, allowing for a systematic analysis of idle time and the optimization of critical paths of computation. Moreover, as algorithm-level descriptions are translated into compiled circuits, hardware-specific execution characteristics become crucial in determining performance. Integrating these factors at the compilation stage can lead to more efficient implementations tailored to specific quantum architectures.

In this work, we enable the concurrent execution of operations that were previously sequential by strategically reordering midlevel computational tasks, leading to enhanced performance. As part of this approach, we introduce new designs for both the standard and discrete logarithm variant of Shor's factoring algorithm that reduce idle time while maintaining qubit efficiency. We extend these methods to distributed quantum computing (DQC) [16], [17], [18], [19], [20], demonstrating how task rearrangements optimize execution when multiple ebit channels are available. To systematically assess the impact of different design choices, we utilize static timing analysis (STA) [21], [22], a technique widely applied in classical circuit design to analyze execution timing under hardware constraints. This allows us to assess circuit delays across various quantum hardware platforms and gain insights into how execution bottlenecks arise at different levels of abstraction.

Our results indicate that the proposed task parallelization methods remain effective regardless of the target hardware architecture. In monolithic systems, we demonstrate that idle time can be significantly reduced and quantify the tradeoffs between qubit efficiency and execution time, particularly in architectures with slow reset and measurement operations. In distributed setups, our analysis reveals how execution time is influenced by the interplay between ebit generation time, the number of ebit channels, and the size of the factored number. To evaluate our approach, we integrate modular exponentiation circuits from the QRISP library [23] into our designs, enabling us to construct circuits capable of factoring large numbers. We conduct a comprehensive study across neutral atom, superconducting, and ion-trap quantum computing platforms, analyzing how our designs impact circuit delays. The insights gained from this analysis provide a starting point for optimizing quantum circuit compilation based on specific hardware constraints.

The rest of this article is organized as follows. Section II introduces Shor's algorithm, key concepts in dynamic circuits, DQC, and the STA techniques used for execution time analysis. Section III presents our midlevel abstraction perspective, examines existing circuit designs in terms of qubit count and idle time, introduces our proposed circuit designs, and extends this perspective to distributed execution. Section IV details our evaluation methods, including circuit construction using QRISP, hardware modeling, and delay estimation. Section V presents our experimental results and their implications for quantum circuit design. Finally, Section VI concludes this article and presents a discussion of future research directions.

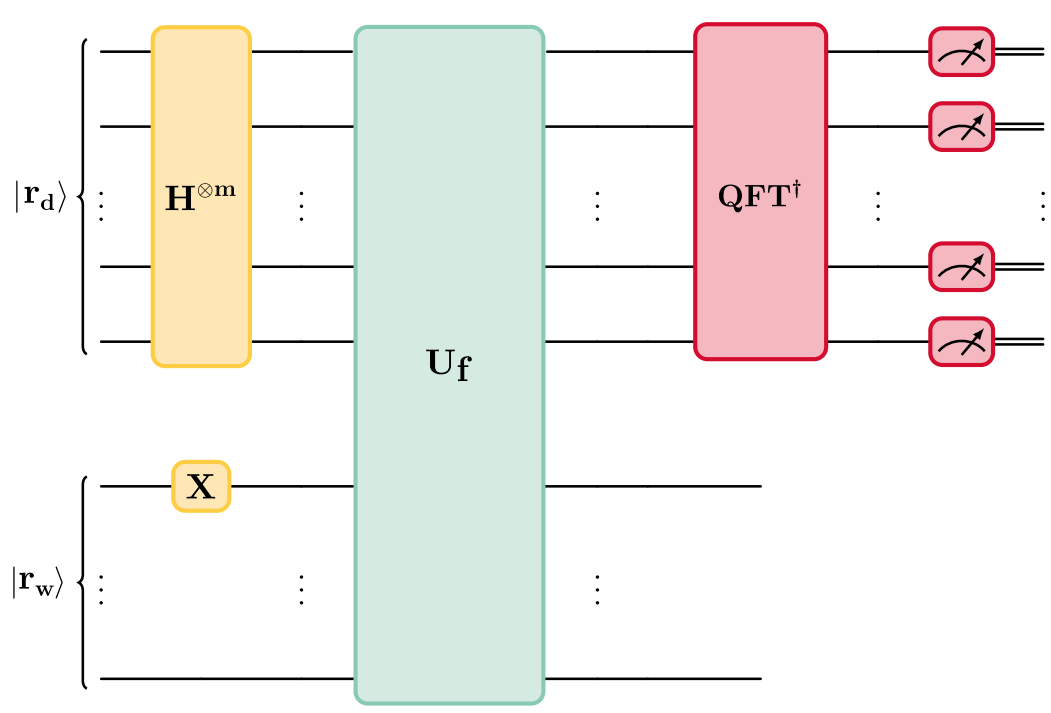

**FIGURE 1. High-level circuit diagram of Shor's algorithm for prime factorization.**

## II. BACKGROUND

### A. SHOR'S ALGORITHM

Shor's algorithm [1] is designed to factor a large composite integer $N$ into its prime factors $p$ and $q$ where $N = p \cdot q$.[1] The factorization problem is reformulated as an order-finding problem. The goal is to find the period $r$ of the modular exponentiation function $f(x) = a^x \mod N$, for a randomly chosen positive integer $a < N$ coprime to $N$. Once $r$ is obtained, the unknown factors $p$ and $q$ can be computed classically by solving $\gcd(a^{\frac{r}{2}} \pm 1, N)$ through the Euclidean algorithm. By proposing an efficient quantum subroutine for finding $r$, Shor's algorithm can solve the factoring problem in polynomial time on a quantum processing unit (QPU).

The quantum-order finding subroutine relies on the unitary implementation of the modular exponentiation function $f(x)$

$$U_f : |x\rangle\,|y\rangle \mapsto |x\rangle\,|y \cdot a^x \bmod N\rangle\,. \tag{1}$$

Here, the register $|y\rangle$, used for computing the modular exponentiation, is called the *work register* (denoted as $|r_w\rangle$) and requires $n$ qubits, where $n = \lceil \log_2 N \rceil$. The data register (denoted as $|r_d\rangle$), which stores the exponent $x$, has a size of $m = 2n$.

The algorithm consists of three steps: 1) initialize the work register $|r_w\rangle$ to 1 and the data register $|r_d\rangle$ to an equal superposition using Hadamard gates; 2) apply the unitary operation $U_f$; and 3) apply the inverse quantum Fourier transform (QFT$^\dagger$) to the data register $|r_d\rangle$. Upon measuring the final state, an $m$-bit estimate of $j/r$ is obtained, where $j \in \{0, \ldots, r-1\}$. The order $r$ can then be determined through classical postprocessing using continued fractions. Fig. 1 shows the complete high-level circuit of Shor's algorithm.

While this first variant of Shor's algorithm is most commonly studied, a second variant of using Shor's algorithm for factoring has been introduced that connects order finding to the problem of computing discrete logarithms [25], [26]. In this case, for $a$ coprime to $N$ and $b = a^d \mod N$, the goal is to compute the exponent $d = \log_a b$. Shor's algorithm solves this problem by constructing the bivariate function

$$f(x_1, x_2) = a^{x_1} b^{-x_2} \mod N \tag{2}$$

and using order finding to identify the order $(d, 1)$. For the order $d = \log_a b$, we know that $d = p + q$ if $r > p + q$, which holds with a large probability for sufficiently large $N$. Since $N = pq$, if we can compute $d = p + q$, we can solve the simple quadratic equation $p^2 - dp + N = 0$ to recover $p$ and $q$. In this implementation of Shor's algorithm, two data registers $|x_1\rangle$ and $|x_2\rangle$ are needed to store the two exponents. This variant has the following advantage: the value $d$ to be recovered is typically smaller than the period $r$ from the standard variant, which reduces the required size of the data registers. Even though we now need two data registers, we can choose them of size $2m$ for $x_1$, and $m$ for $x_2$, where $m = 0.5n + O(1)$. This gives a combined data register size of only $3m = 1.5n + O(1)$.

[1] For a more detailed introduction, we refer to [24, Sect. 5].

The discrete logarithm variant was used in the most established work for realistic requirements on factoring numbers in the order of magnitude used by modern encryption protocols [5]. This was recently updated [27] by incorporating improved adder circuits [28] and logical arithmetic techniques.

The efficiency of both variants of Shor's algorithm primarily depends on the implementations of the quantum Fourier transform (QFT) and modular exponentiation operations $U_f$. For the QFT, a well-established implementation exists using a gateset composed of controlled phase rotations and Hadamard gates. However, the primary bottleneck lies in the efficient implementation of the modular exponentiation $U_f$, which poses a substantial challenge. Since $U_f$ must be unitary, reversible arithmetic techniques, such as uncomputation [6], must be employed. In addition, extra ancilla qubits are often required for the work register.

At the lowest level of abstraction, the core building block for modular exponentiation is a modular adder. Various types of adders have been proposed [7], [8], [9], [10], [11], [12], [13] and can be classified based on their underlying approaches, including ripple carry [7], [8], carry lookahead [9], and QFT-based methods [10], [12], [13]. The modular operations are then built hierarchically—first performing modular addition, then modular multiplication, and finally modular exponentiation. Implementations are typically evaluated based on tradeoffs between circuit width and depth. For instance, some approaches minimize the required qubit count by increasing circuit depth [13], [14], [15], achieving a current minimum of $2n + 2$ qubits through the reuse of dirty ancillas [14], [15].

Specialized circuits have also been developed to demonstrate Shor's algorithm on noisy intermediate-scale quantum (NISQ) hardware [29], [30], [31]. These circuits are often optimized for specific values of $N$ and $a$, rather than being general-purpose solutions. For detailed discussions on optimization techniques, such as window arithmetic, coset

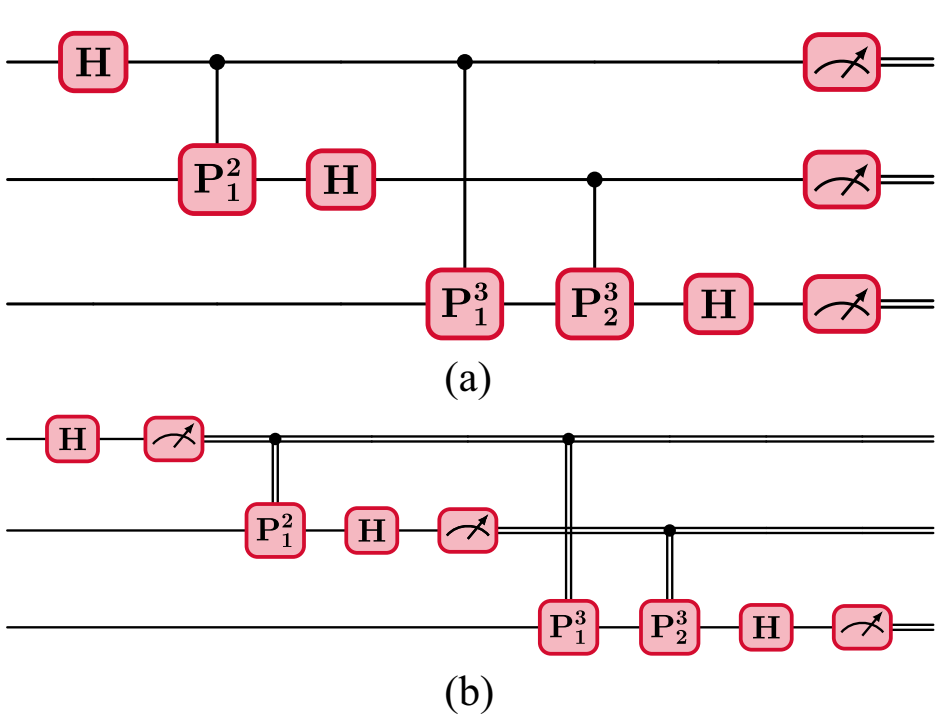

**FIGURE 2.** **(a) Standard QFT implementation utilizing two-qubit controlled rotations. (b) Semiclassical QFT using early measurements to replace the two-qubit gates with classically controlled single-qubit phase rotations.**

representations, and comparisons of adder implementations, we refer readers to [5] and [10].

Our proposed designs operate at a midlevel abstraction layer. We consider the variants of Shor's algorithm discussed earlier and decompose the components into tasks. The midlevel abstraction is based on the interpretation of the algorithm as an application case of quantum phase estimation (QPE), and without relying on specific arithmetic implementations. This approach ensures compatibility with a wide range of low-level optimizations, including various modular addition realizations.

### *B. DYNAMIC CIRCUITS*

Circuits that use midcircuit measurements, qubit resets, and feedforward control based on classical information are referred to as dynamic circuits [32], [33], [34]. These circuit elements are not only essential for implementing quantum error detection and correction [35], [36], but have also recently been found to aid in other tasks, such as enabling efficient long-range entanglement [37], [38] and state preparation [39], [40].

Dynamic circuits play a key role in optimizing Shor's algorithm. This is especially relevant due to the introduction of the semiclassical implementation of the QFT [41], [42]. If the QFT is followed by measurement rather than additional gates, the principle of deferred measurement [24] can be applied. This enables earlier measurements and substitutes each two-qubit controlled rotation with a single-qubit rotation, governed by a classical measurement result, in the QFT realization (see Fig. 2). In iterative phase estimation (IPE) [43], [44], the semiclassical QFT is used within QPE, making it possible for each qubit of the QFT to be executed sequentially. Since no multiqubit gates are involved, a single qubit is sufficient if qubit resets are available. Shor's order-finding routine, which—both in its standard and discrete logarithm variant—can be viewed as a form of QPE, benefits from IPE to reduce the required qubit count [45], [46]. This iterative approach has been experimentally demonstrated for two-digit values of $N$ [47].

**FIGURE 3.** Distributed setup with two QPUs, *A* and *B*, linked by an ebit channel that generates ebits $|\Phi^+\rangle_{AB}$ on communication qubits $c_1$ and $c_2$, facilitates computation involving computing qubits $qi$ and $qj$, located on separate QPUs.

Finally, even though dynamic circuits are now supported by several hardware architectures (including superconducting [34], ion-trap [36], and neutral atom platforms [48], [49]), finding practical use cases where these tools can be effectively utilized remains a challenge. As of now, making use of dynamic circuits is not done systematically and often requires a careful case-by-case inspection of circuit structure.

### C. DISTRIBUTED QUANTUM COMPUTING

Current quantum hardware faces limitations in qubit scaling due to technology-specific engineering challenges. Superconducting quantum computers rely on dilution refrigerators to keep qubits at low temperatures. As the number of qubits increases, the complexity and cost of managing wiring, control systems, and refrigeration also rise significantly [50]. In neutral atom systems, efficiently scaling qubit arrays requires high-intensity lasers to maintain strong optical tweezers while optimizing spatial constraints [51]. Trapped-ion systems face challenges with frequency crowding, where closely spaced motional modes complicate individual qubit addressing without crosstalk, necessitating precise control lasers [52]. To address these challenges and enable interaction between distant quantum computers, DQC has been proposed. DQC involves operating multiple QPUs interconnected in a network, enabling scalable quantum processing [16]. This section offers an overview of the key terms and concepts from [17] relevant to this work.

In a DQC environment, each QPU contains a set of qubits, which are classified into two types: compute qubits, used for general computations, and communication qubits. The communication qubits are responsible for generating shared entanglement between distant QPUs. DQC protocols depend on communication *ebits* sharing the Bell state

$$|\Phi^+\rangle_{AB} = \frac{1}{\sqrt{2}}(|0\rangle_A\,|0\rangle_B + |1\rangle_A\,|1\rangle_B) \tag{3}$$

where the first qubit from the pair helps communication with the compute qubits on QPU $A$, while the second qubit helps communication with the compute qubits on QPU $B$.

A communication channel where such ebits sharing the Bell state $|\Phi^+\rangle_{AB}$ are repeatedly generated is called an ebit channel (see Fig. 3).

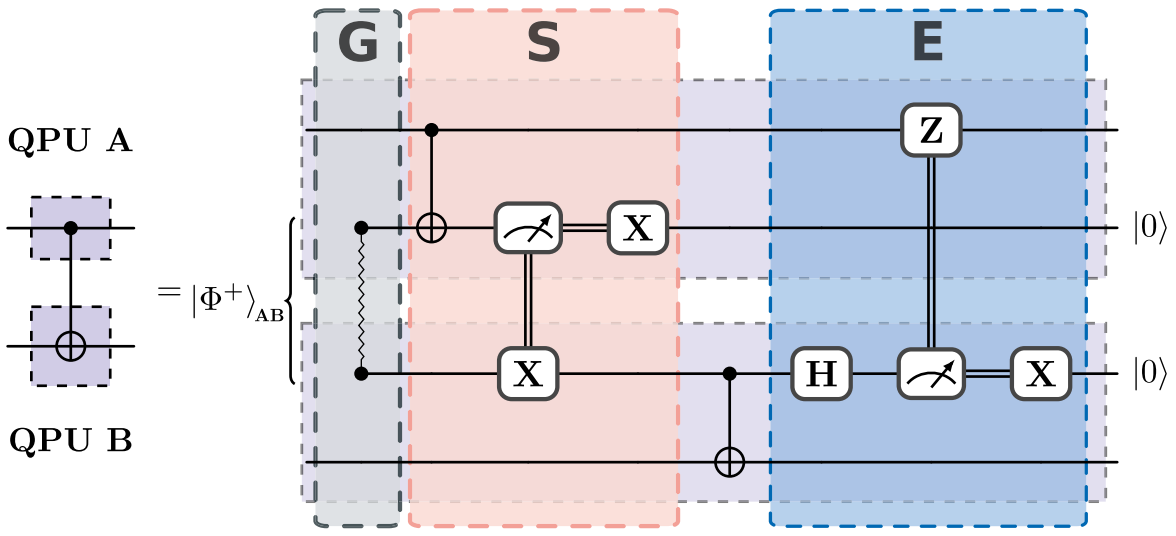

**FIGURE 4.** Steps involved in conducting a remote CNOT operation between QPUs *A* and *B* using the EJPP protocol: ebit generation (G), the EJPP starting phase (S), local execution of the CNOT gate, and the EJPP ending phase (E).

Analogous to classical communication, when computing nodes lack direct ebit channels, quantum repeaters, and quantum routers leveraging the entanglement swapping (ES) protocol have also been proposed [18], [53], [54] to facilitate communication.

There are two primary protocols for utilizing ebits in distributed computation. The first is the teleportation or teledata protocol [55], which consumes one ebit to transfer the state of a qubit from one QPU to another.

The second is the EJPP protocol, also known as telegate [56]. This protocol enables the remote execution of certain gates, with controlled gates being the most notable example.

The EJPP protocol is executed in three phases, as shown in Fig. 4: Initially, during the starting process (also known as cat-entangler phase), the state of the compute qubit from one QPU is temporarily mirrored onto the communication qubit of the ebit associated with the compute qubit on the other QPU. Then, a controlled gate is executed locally between the mirrored communication qubit and the target compute qubit. Finally, in the ending process (also called the cat-disentangler phase), the state of the control compute qubit is decoupled from the communication qubit.

The EJPP protocol gives an advantage over teledata when executing multiple controlled gates using a single ebit, provided that no interrupting conditions are met. Such conditions include an intermediate Hadamard gate on the control qubit of two CNOT gates or a pair of CNOT gates acting on the same qubits with flipped control and target. As a result, EJPP enables remote execution of a $CU$ operation involving a complex multiqubit $U$ with just one ebit, even if $U$ is decomposed into multiple simpler controlled gates.

### D. STATIC TIMING ANALYSIS

Currently, the time analysis of quantum circuits is primarily approached from the standpoint of computational complexity, i.e., the scaling of algorithms in terms of O-notation [57]. However, a more practical approach to time analysis is becoming essential as quantum hardware advances. The supported gate set and corresponding gate times vary depending on the targeted hardware. In addition, factors like the execution time of measurement and reset operations, as well as qubit coherence times, differ depending on the chosen QPU. To assess practical feasibility and optimize circuits effectively, it is important to account for these specific execution times. Some works already consider such hardware timing aspects in the context of circuit cutting [58], circuit compilation [59], and qubit mapping [60], [61], [62]. With the progress in quantum error correction, timing will become an even more critical consideration.

**TABLE 1. Abbreviations and Adopted Notation From STA [21], [22]**

| Term/Symbol | Definition |
|---|---|
| $N$ | Number to be factored |
| $n$ | Number of bits of $N$, i.e., $\lceil \log_2 N \rceil$ |
| Gate delay $t_U$ | Time for an input state to transition a gate $U$ |
| Circuit graph | Directed acyclic graph (DAG) modeling gate dependencies |
| Weighted circuit graph | Weighted directed acyclic graph (WDAG) extending the DAG with a weight function representing gate delays |
| Critical path | Path with longest delay, determining the delay of the overall circuit |
| Circuit depth | Length of critical path in the circuit DAG |
| Path delay $t_P$ | Combined signal delay along a path $P$ of gates in the WDAG |
| Circuit delay $t_C$ | Path delay of critical path |

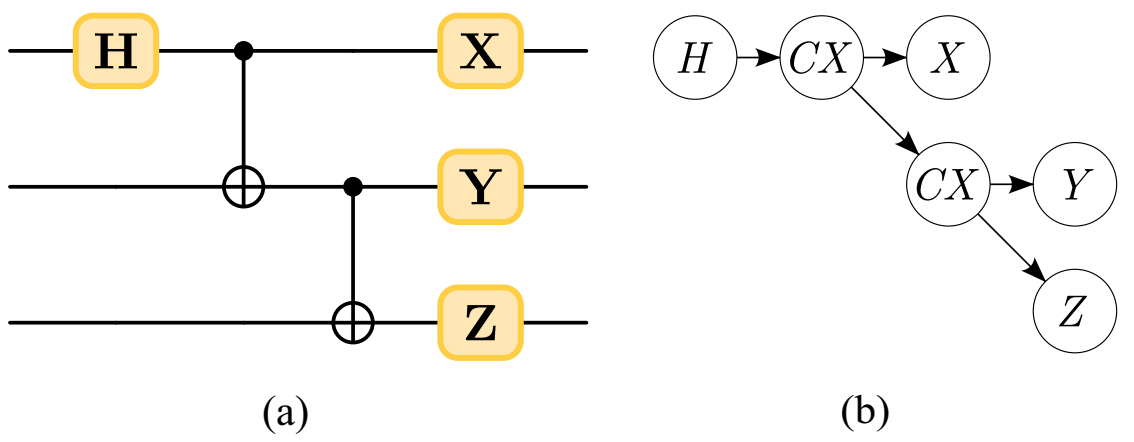

**FIGURE 5. (a) Example quantum circuit. (b) Corresponding DAG representation. The DAG has two longest paths: *H* ⇝ *CU* ⇝ *CU* ⇝ *Y* and *H* ⇝ *CU* ⇝ *CU* ⇝ *Z*.**

Furthermore, in the context of DQC, the time required for ebit distribution and routing through quantum networks must also be considered [63], [64], [65], [66], [67]. As quantum systems become larger and more distributed, resources for distribution and the scheduling and synchronization of computing tasks (e.g., with an execution manager [68]) will be necessary. Therefore, simply analyzing the runtime of a circuit from a complexity standpoint is no longer sufficient. This challenge is not new: classical circuits are also evaluated based on their theoretical complexity but must be examined with specific timing constraints for practical circuit design [21], [22]. We adopt definitions of classical circuit design automation, similar to how definitions of classical circuit complexity are used in quantum circuit complexity [57]. A summary of definitions and notation is given in Table 1.

*Definition 1 (Gate Delay):* The gate delay is defined as the time required to transform the state of $n$ qubits (where $n \geq 1$) when applying an $n$-qubit gate $U$ and denoted as $t_U$.

In classical circuits, the delay of a logical gate refers to the time it takes for a signal to pass through it. In quantum circuits, we define such delay as the gate time of the native gate set supported by the targeted QPU.

*Definition 2 (Circuit Graph):* A circuit graph is a directed acyclic graph (DAG), where each vertex $v \in V$ represents a gate, and an edge $(u, v) \in E$ indicates that gate $v$ is a direct successor of gate $u$ in a given quantum circuit $C$, and denoted as $G_C = (V, E)$. The set $V$ also includes two additional vertices, $Sc$ and $Sk$, which have no predecessor and successor nodes, respectively, to mark the start and end of the circuit execution.

Fig. 5 illustrates an example circuit along with its corresponding DAG. The depth of a circuit is typically represented as the number of distinct time steps required to execute all circuit instructions. In the DAG representation, this can be defined more precisely.

*Definition 3 (Critical Path, Circuit Depth):* The critical path is the longest sequence of vertices of the form $Sc \rightsquigarrow Sk$ that are connected by edges for a given circuit graph $G_C$ and the length of the critical path is referred to as the circuit depth. The critical path is not necessarily unique.

This aligns with the intuitive definition, as each vertex represents an instruction and each edge indicates a direct dependence. Hence, all instructions on critical paths must be executed sequentially in separate time steps. Since the instructions on critical paths determine the overall execution time of the circuit, they act as a bottleneck in the computation. Understanding how the structure of a circuit defines these critical paths can provide valuable insights for optimization.

*Definition 4 (Weighted Circuit Graph):* The weighted circuit graph is a circuit graph $G_C(V, E)$, with edges like $(u, v)$ weighted according to the gate delay $t_v$ of vertex $v$ and denoted by $G_W(V, E, w)$. For all edges like $(u, Sk)$ where $Sk$ indicates the end of execution, the weight $w$ is set to 0.

The WDAG not only represents the dependencies between gates but also incorporates their associated delays. Fig. 6 illustrates an abstract Shor circuit along with its corresponding DAG and WDAG.

*Definition 5 (Path Delay):* The path delay is the sum of weights of the edges along a given path $P$ in a weighted circuit graph $G_W(V, E, w)$ and denoted as $t_P$.

*Definition 6 (Circuit Delay):* The circuit delay is the critical path delay in a weighted circuit graph $G_W(V, E, w)$ and denoted as $t_C$. For a sequence of operations $U_1, \ldots, U_k$ forming a circuit $C$, we also denote the circuit delay $t_C = t(U_1 U_2 \ldots U_k)$. The circuit delay is bound by the sum of the delay of all its operations, i.e., $t_C \leq \sum_{i=1}^{k} t_{U_i}$, where the inequality becomes an equality only if the $U_i$ operations form a path.

Circuit delay, like circuit depth, measures a circuit's execution time but with greater accuracy for specific hardware by considering gate execution times. This is important because gate delays influence path delays and can alter the longest path within the circuit. Consequently, the critical path

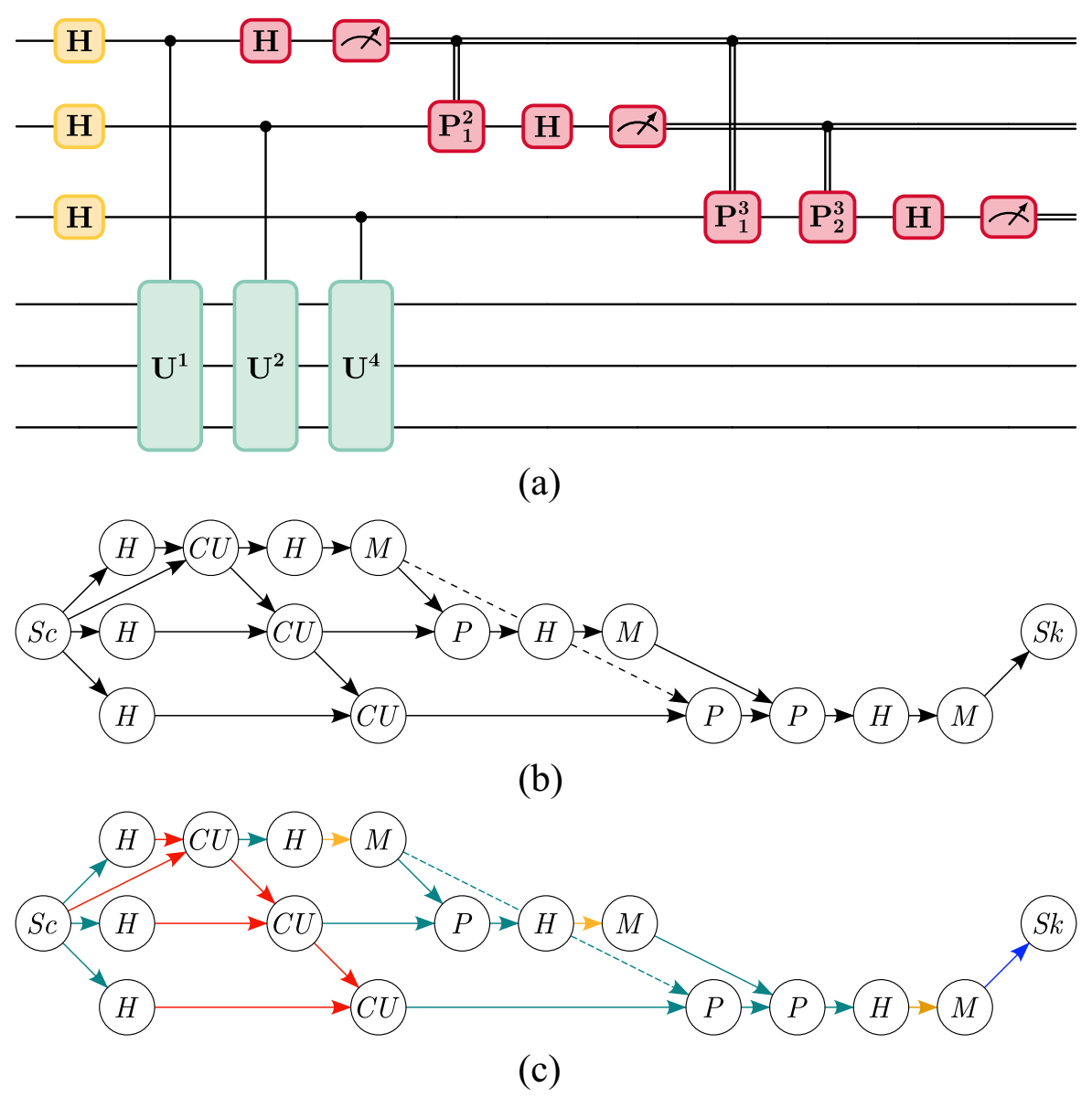

**FIGURE 6.** (a) Abstract Shor circuit and corresponding, (b) DAG, and (c) WDAG representation. The colors of incoming edges in the WDAG represent the magnitude of gate delays. (Light blue) Single-qubit gates have short delays, (yellow) compared to the medium delay measurements, (red) or time-intensive composite *CU* operations. (Dark blue) Incoming edges to the artificial sink node $S_k$ that indicate the end of computation have no delay.

identified in the DAG representation may not align with that of the WDAG.

## III. PARALLELIZATION OF SHOR'S ALGORITHM

### A. TASK ABSTRACTION

The quantum order-finding subroutine of the standard variant of Shor's algorithm can be viewed from the perspective of QPE. Specifically, consider an $n$-qubit unitary operator $U$ with eigenvalues $\lambda_j = e^{2\pi i\theta_j}$ and corresponding eigenstates $|\psi_j\rangle$. The objective of QPE is to estimate $\theta_j$ given an eigenstate $|\psi_j\rangle$. Since $|\psi_j\rangle$ is an eigenstate, it satisfies the relation $U|\psi_j\rangle = \lambda_j|\psi_j\rangle$. By employing a controlled-$U$ ($CU$) gate and preparing the control register in the state $|+\rangle = 1/\sqrt{2}(|0\rangle + |1\rangle)$, the eigenvalue can be moved to the control register using phase kickback

$$CU(|+\rangle|\psi_j\rangle) = \frac{1}{\sqrt{2}}(|0\rangle|\psi_j\rangle + \lambda_j|1\rangle|\psi_j\rangle). \tag{4}$$

By applying a Hadamard gate to the control register and then measuring it, we can extract $\lambda_j$ and consequently determine $\theta_j$, provided $\lambda_j \in \{\pm 1\}$. For higher resolutions of $\theta_j$, i.e., $\theta_j = \frac{k}{2^m}$, the approach can be extended by using an $m$-qubit control register $|l\rangle$ and replacing the $CU$ gate with an $m$-controlled-U ($C_mU$) gate such that

$$C_mU : |l\rangle|\psi_j\rangle \mapsto |l\rangle U^l|\psi_j\rangle. \tag{5}$$

The desired value of $\theta_j$ is obtained by performing QFT$^\dagger$ on the control register, followed by measurement. The general structure of the algorithm is depicted in Fig. 7.

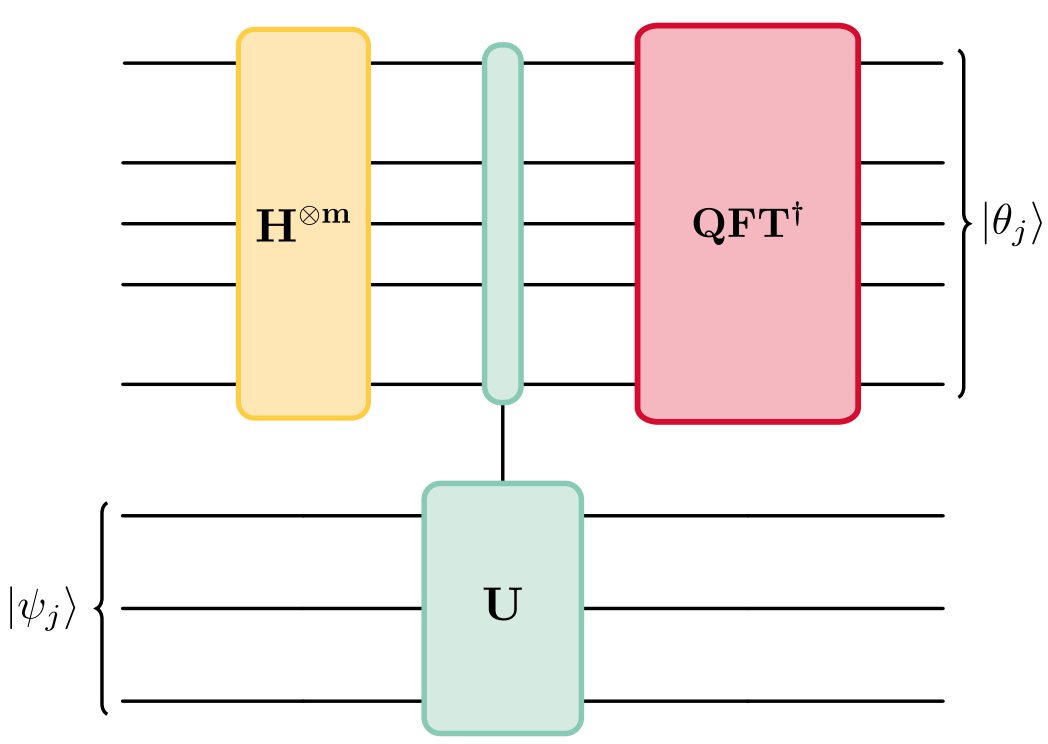

**FIGURE 7.** High-level circuit diagram of QPE. Applying QPE on the eigenstate $|\psi_j\rangle$ yields the phase $\theta_j$ of eigenvalue $\lambda_j = e^{2\pi i\theta_j}$.

When using Shor's algorithm for solving the order-finding problem to factor a number $N$, the circuit can be adapted by choosing the $U$ operation as

$$U_a : |x\rangle \mapsto |x \cdot a \mod N\rangle. \tag{6}$$

Then, $C_mU_a$ is equivalent to $U_f$ from (1). In this setup, the $m$-qubit control register $|l\rangle$ corresponds to the data register $|r_d\rangle$, while the bottom register holding the eigenstate $|\psi_j\rangle$ corresponds to the work register $|r_w\rangle$.

Furthermore, the state preparation of $|\psi_j\rangle$ can be avoided in this circuit. This is because the state $|1\rangle$ is an equal superposition over all eigenstates of $U_a$. To extract the order $r$, any $\theta_j$ where $j$ is coprime to $r$ can be used. Thus, initializing the work register in the state $|1\rangle$ and running the algorithm effectively results in a random selection of one of the eigenstates.

Our circuit optimization schemes do not require decomposing $U_a$ into elementary gates. The designs make use of the inherent structure of QPE and can, in principle, be applied to any QPE-based application. Specifically, in the context of Shor's algorithm, these optimizations can be integrated with any modular adder circuit or other enhancements, as long as they do not interfere with the circuit's higher-level abstraction.

The $C_mU$ operation from (5) can be decomposed as follows: Since $l = \sum_{i=0}^{n-1} l_i 2^i$ is represented in binary on the data register, each bit $l_i$ accounts for controlling $2^i$ repetitions of $U$, i.e., a $CU^{2^i}$ operation. Therefore, the $C_mU$ operation can be expressed as a sequence of $m$ gates acting on the work register, which is initialized in the state $|1\rangle$. Each $CU^{2^i}$ gate is controlled by a separate qubit from the data register $|r_d\rangle$, as shown below

$$C_mU(|l\rangle|\psi\rangle) = \prod_{i=0}^{n-1} CU^{2^i}(|l\rangle|\psi\rangle). \tag{7}$$

The QFT$^\dagger$ has a sequential structure. For each qubit $q_j$, a phase correction operation $P_j$ is applied, followed by a Hadamard gate. The phase correction operation $P_j$ on qubit $q_j$ involves a set of controlled phase rotations $R_k^{-1} = P(-2\pi i/2^k)$ determined by the states of the preceding qubits,

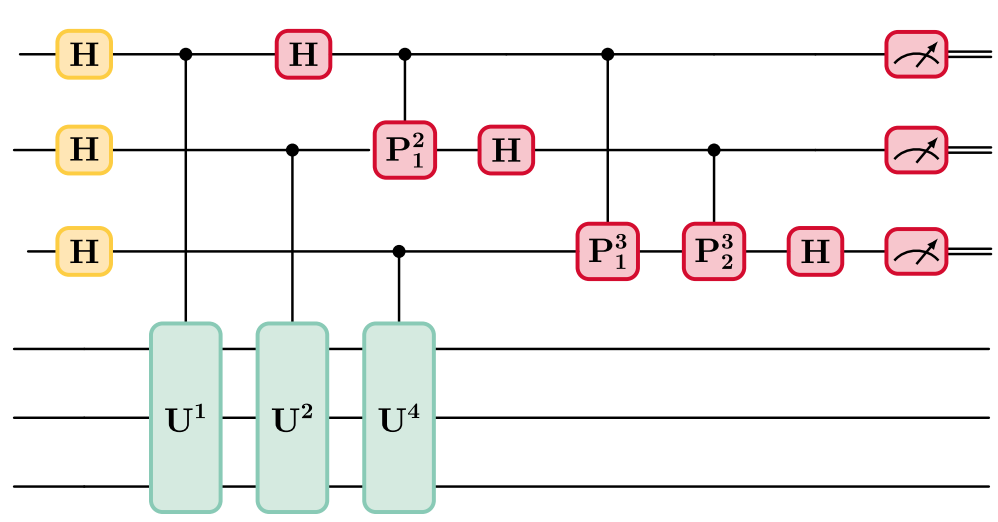

**FIGURE 8.** High-level circuit for the regular design of Shor's algorithm, implemented as QPE with a three-qubit data register (i.e., $m = 3$). Each qubit in the data register acts as a control qubit for a specific modular exponentiation operation of the form $U^{2^i}$, followed by phase processing as part of the QFT$^{\dagger}$.

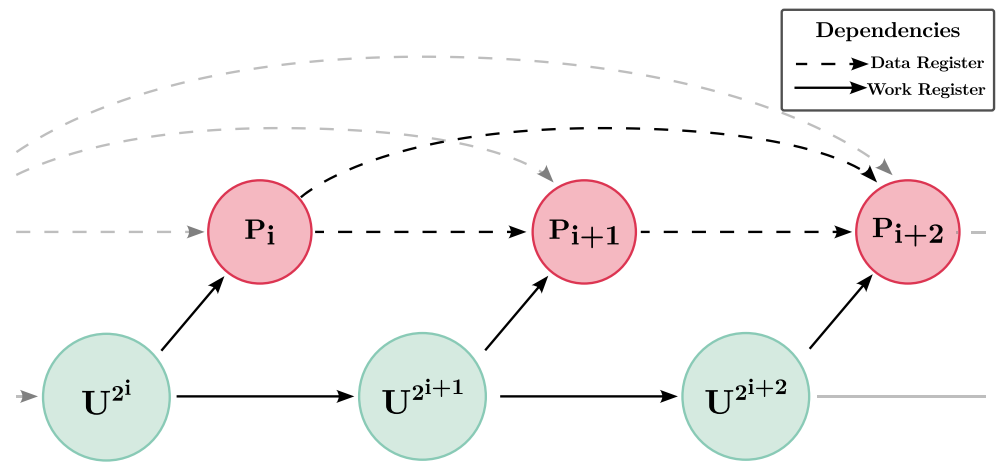

**FIGURE 9.** Dependencies of high-level tasks of Shor's algorithm in DAG form. Each $CU^{2^{i+1}}$ operation depends on its predecessor $CU^{2^i}$ on the work register, and each phase processing $P_{i+1}$ depends on its corresponding $CU^{2^{i+1}}$ operation on the work register and all previous $P_i \ldots P_1$ on the data register.

specifically

$$P_j = \prod_{k=1}^{j} R_{k+1}^{j-k} \tag{8}$$

where $R_k^l$ is the phase rotation $R_k$ controlled by qubit $q_l$. Applying QFT$^{\dagger}$ to the data register $|r_d\rangle$, we obtain

$$\text{QFT}^{\dagger}\,|r_d\rangle = \prod_{i=0}^{m-1} P_i H_i\,|r_d\rangle \tag{9}$$

where $H_i$ denotes the Hadamard gate applied to qubit $q_i$. This standard approach is referred to as the *regular* design, and Fig. 8 shows an exemplary circuit for $m = 3$.

To summarize, the main operations for the work register involve the $m$ $CU^{2^i}$ gates, which are executed sequentially. Each qubit in the data register controls exactly one of these gates and then undergoes a phase correction step depending on all preceding data qubits, as shown in Fig. 9.

For the discrete logarithm variant, the core structure of the algorithm stays the same. Instead of one data register $|r_d\rangle$, we now have two data registers $|r_{d_1}\rangle$ , $|r_{d_2}\rangle$, and two different types of $U$ operations. $|r_{d_1}\rangle$ controls $U_a$ operations according to (6) and $|r_{d_2}\rangle$ controls a similar operation $U_{b^{-1}}$, but with a different precomputed constant $b^{-1}$, where $b = a^{N+1}$:

$$\begin{aligned} C_m U_{b^{-1}} C_{2m} U_a\,|x_1\rangle\,|x_2\rangle\,|\psi_j\rangle &= C_m U_{b^{-1}}\,|x_1\rangle\,|x_2\rangle\,U_a^{x_1}\,|\psi_j\rangle \\ &= |x_1\rangle\,|x_2\rangle\,U_{b^{-1}}^{x_2} U_a^{x_1}\,|\psi_j\rangle\,. \end{aligned} \tag{10}$$

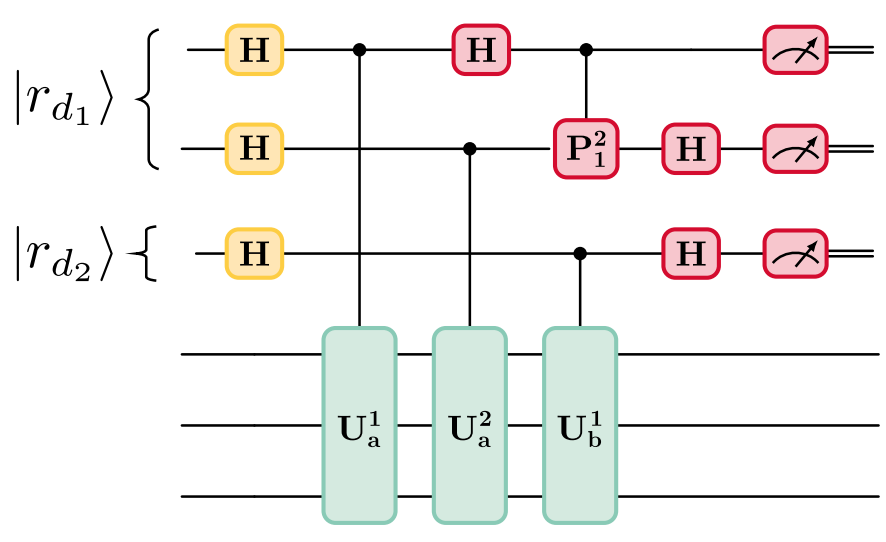

**FIGURE 10.** High-level circuit for the regular design of the discrete logarithm variant of Shor's algorithm, implemented as QPE with a two-qubit $|r_{d_1}\rangle$ and a one-qubit $|r_{d_2}\rangle$ data register. Each data qubit controls the execution of a specific modular exponentiation operation of the form $U^{2^i}$, followed by phase processing as part of the two separate QFT$^{\dagger}$.

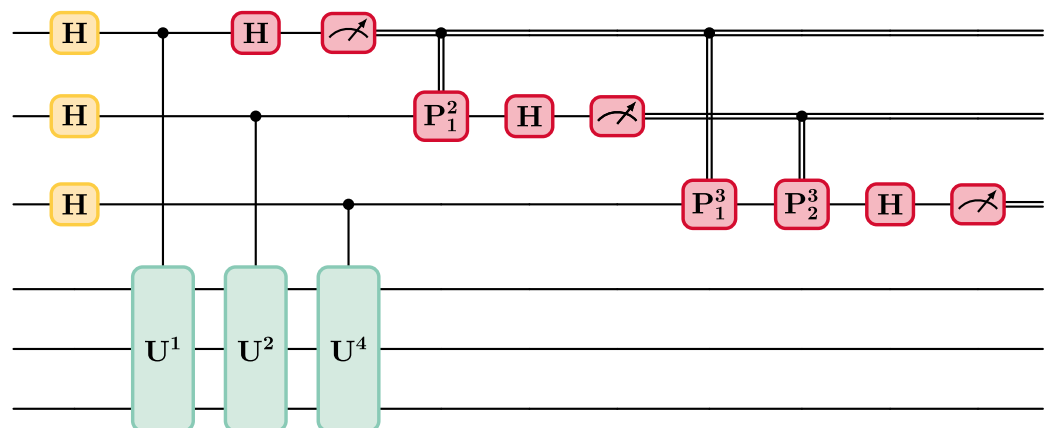

**FIGURE 11.** Regular design of Shor's algorithm utilizing the semiclassical QFT$^{\dagger}$. All two-qubit gates in QFT$^{\dagger}$ acting on three-qubit data register (i.e., $m = 3$) are replaced with classically controlled single-qubit gates.

The desired multiproduct $f(x_1, x_2)$ from (2) is then realized on the work register by again initializing the work register $|r_w\rangle$ to $|1\rangle$ and applying the two $U$ operations

$$\begin{aligned} U_{b^{-1}}^{x_2} U_a^{x_1}\,|1\rangle &= U_{b^{-1}}^{x_2}\,|a^{x_1} \mod N\rangle \\ &= |(b^{-1})^{x_2}(a^{x_1} \mod N) \mod N\rangle \\ &= |a^{x_1} b^{-x_2} \mod N\rangle \\ &= |f(x_1, x_2)\rangle\,. \end{aligned} \tag{11}$$

Initializing the two work registers with Hadamard gates then applies $f(x_1, x_2)$ on the work register with an equal superposition of exponents. Finally, two separate QFT$^{\dagger}$ are applied to $|r_{d_1}\rangle$ and $|r_{d_2}\rangle$ after the $U$ operations are finished.

Since the $C_m U$ construction is the same as in the standard variant, we can similarly decompose both $C_m U_{b^{-1}}$ and $C_{2m} U_a$ by interpreting the data registers in their binary form. An example of the regular design for the discrete logarithm variant is shown in Fig. 10.

### B. PARALLELIZATION IN MONOLITHIC ARCHITECTURE

An effective optimization technique to reduce the qubit count in Shor's algorithm is to use the semiclassical implementation of the QFT [41]. This approach leverages the principle of deferred measurement [24]. Specifically, after applying the final Hadamard gate to a qubit, the qubit can be measured immediately, as all later operations are controlled gates that can be implemented using classical controls. This process is illustrated in Fig. 11.

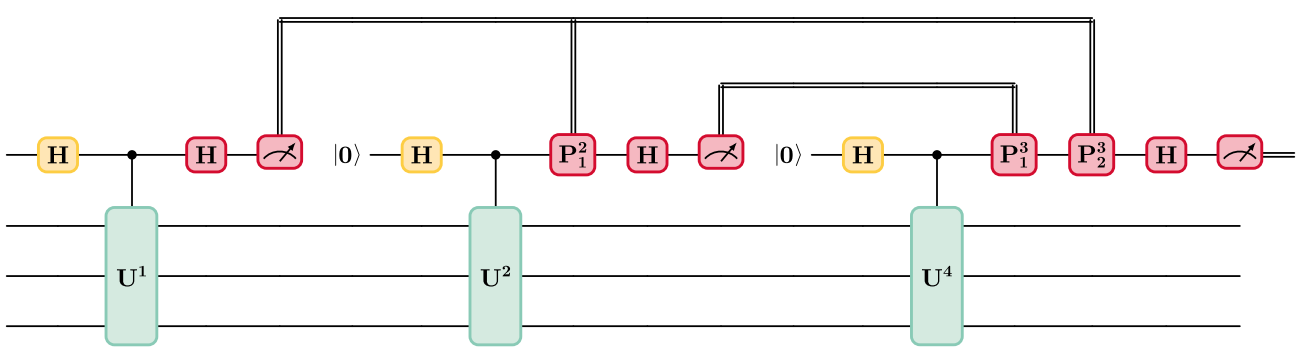

**FIGURE 12.** Iterative design of Shor's algorithm with a three-qubit data register (i.e., $m = 3$) implemented as IPE. The data register is reduced to a single qubit at the expense of longer circuit delay.

The IPE [43], [44] incorporates the semiclassical QFT within QPE. In this approach, once a data qubit has been used as a control for its corresponding $CU^{2^i}$ operation, completed its phase processing, and its phase has been measured, it is no longer needed. This allows the qubit to be reset and reused for the next $CU^{2^i}$ operation. The sequence of operations on the $i$th qubit then forms an iterative package of operations $I_i$

$$I_i = RMHP_iCU^{2^i}H. \tag{12}$$

Here, the index of the initial Hadamard gate $H$ is omitted since only one qubit is used for the data register. $M$ denotes measurement in the computational basis, $R$ represents the qubit reset, and $P_i$ corresponds to the phase correction described in (8). As each package $I_i$ is independent of the data qubit of the previous package $I_{i-1}$ and only requires its classically stored measurement result, the entire data register can be implemented using just a single qubit. All iterative packages $\{I_i\}_{i=0}^{m-1}$ can simply be processed sequentially on a single data qubit, which can be represented as follows:

$$\prod_{i=0}^{m-1} I_i. \tag{13}$$

This method is known as the *iterative* realization of Shor's algorithm. An example circuit for the case $m = 3$ is shown in Fig. 12.

Although the iterative realization uses only one qubit instead of $m$ qubits for the data register, it results in a longer execution time. As outlined in (13), each operation must be performed sequentially: starting with the Hadamard initialization, followed by the controlled modular arithmetic $CU^{2^i}$, phase correction $P_i$, a second Hadamard gate, measurement, and finally a qubit reset. This sequential execution creates a critical path that traverses all $CU^{2^i}$ operations for $i = 0, 1, \ldots, m-1$, along with the phase correction steps. Throughout this process, the work register remains active only during the $CU^{2^i}$ operations, leading to idle periods, as illustrated in Fig. 13.

In contrast, in the regular realization, once a $CU^{2^i}$ operation is completed on the work register, the next $CU^{2^{i+1}}$ operation can proceed without waiting for phase processing, measurements, resets, and reinitialization. Consequently, the regular approach enables parallel execution of computational tasks, reducing the overall processing time, at the cost of more qubits. The dependencies in Fig. 9 illustrate the parallelization possibilities. The $CU$ operations and $P_i$ operations have sequential dependencies, but between the different types of tasks, there are often no direct dependencies (e.g., $CU^{2^{i+1}}$ and $P_i$), and a parallel execution of tasks is possible.

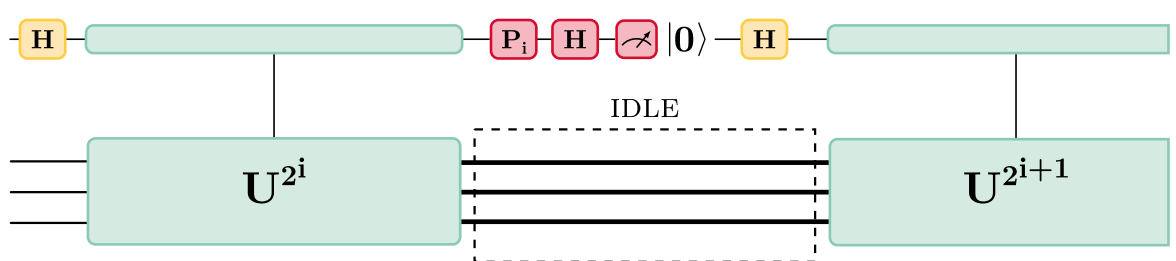

**FIGURE 13.** Idle time experienced in the work register of the iterative design of Shor's algorithm due to phase processing on the data qubit as part of the semiclassical QFT$^\dagger$ operation.

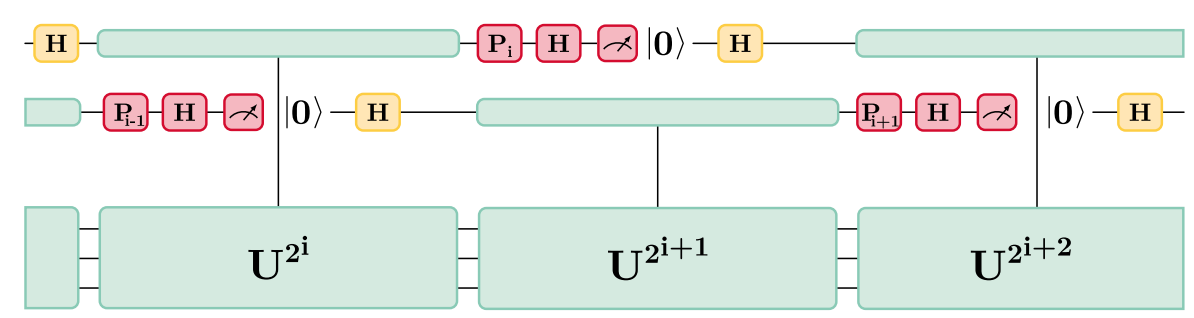

**FIGURE 14.** Alternating design of Shor's algorithm reducing the idle time experienced in the work register of the corresponding iterative design by introducing an additional data qubit to alternate the phase processing tasks.

To address this issue, the alternating design combines qubit resets and the semiclassical QFT$^\dagger$, similar to the iterative approach, but uses two data qubits instead of one to enable parallelization. The key idea is to alternate the iterative process between the two data qubits. After the work register is initialized and the first data qubit undergoes the initial Hadamard gate, i.e., $H_0$, the first $CU^1$ operation is executed using this qubit and the work register. While $CU^1$ is being executed, the second qubit is initialized with $H_1$. Once $CU^1$ is complete, the next operation $CU^2$ is executed using the second qubit and the work register. During $CU^2$, the phase information is extracted from the first qubit, which is then reset and reinitialized with $H_0$ and becomes ready for the next operation $CU^4$ as soon as $CU^2$ finishes. This alternating pattern continues until all $CU^{2^i}$ operations are completed, effectively interleaving the usage of the two data qubits to achieve parallelism and minimize idle time in the work register observed in the corresponding iterative design (see Fig. 13), thereby enhancing overall execution time. Fig. 14 illustrates the design of Shor's algorithm using the proposed alternating approach.

The potential timing advantage of the alternating approach depends on the duration of the $CU^{2^i}$ operations and the phase processing gates. If the $CU^{2^i}$ operations are short, there may not be enough time to simultaneously extract the phase and reset the other data qubit, resulting in some idle time—although shorter than the idle time observed in the iterative design. Conversely, if the $CU^{2^i}$ gates are sufficiently long and satisfy the condition

$$t_{CU^{2^{i+1}}} \geq t(I_i) \quad \forall i \tag{14}$$

then there is no delay in the work register because, during each $CU^{2^i}$ operation, there is sufficient time to extract the phase, reset, and reinitialize the other qubit. Consequently, the critical path consistently runs through the work register without being affected by the data register.

To understand the amount of idle time that can be reduced, both by the alternating and regular design, we first split the overall circuit delay $t_C$ into three parts

$$t_C = t_H + \sum_{i=0}^{m-1} t_{CU^{2^i}} + \delta_P. \quad (15)$$

The initial Hadamard as well as all $CU^{2^i}$ operations have to be executed sequentially, so their delays will always contribute to the overall circuit delay. The additional delay based on phase processing denoted as $\delta_P$ varies based on design choice. We again split $\delta_P$ into two parts

$$\delta_P = \delta_P^M + \delta_P^{\neg M} \quad (16)$$

where $\delta_P^M$ denotes the mitigatable delay and $\delta_P^{\neg M}$ denotes the unavoidable delay due to phase processing. In principle, each phase processing step can be executed in parallel to following $CU^{2^i+1}$ gates. The exception is the last phase processing, which incurs an unavoidable delay since no further $CU^{2^i}$ gates remain for overlap

$$\delta_P^{\neg M} = t(I_{m-1}). \quad (17)$$

The delay that can be mitigated from all other phase processing steps depends on the design choice and the length of the $CU^{2^i}$ operations. If the $CU^{2^i}$ operations are long, the critical path goes through the $CU^{2^i}$ gates and no delay of the phase processing contributes to the circuit delay. However, if the $CU^{2^i}$ operations are instant, all phase processing delay contributes. Therefore, the mitigatable delay due to phase processing is bound by

$$0 \leq \delta_P^M \leq \sum_{i=0}^{m-2} t(I_i). \quad (18)$$

As a more relaxed upper limit, the worst case duration $(m-1)t(I_{m-1}) > \delta_P^M$ can be considered, which uses the worst case $t_{P_{m-1}}$ for phase gates, given that each phase correction $P_i$ requires one more adjustment than its predecessor $P_{i-1}$.

These bounds hold for the alternating as well as the regular design. The specific difference in circuit delay then depends on the durations of the $CU^{2^i}$ operations. Longer delays in $CU^{2^i}$ operation create more opportunities for alternating parallelization while maintaining an overall circuit delay comparable to the regular design. In the worst case scenario, if the $CU^{2^i}$ delay significantly exceeds the upper limit $(m-1)t(P_{m-1}HMRH)$, the idle time becomes negligible. Consequently, the overall circuit delays for all three approaches converge, with a slight preference for the iterative design due to its use of one fewer data qubit compared to the alternating approach.

The discussed designs can similarly be applied to the discrete logarithm variant of Shor's algorithm. The QFT$^\dagger$ on both data registers can be replaced with the semiclassical implementation to enable more flexible task arrangements. Since the multiproduct of $f(x_1, x_2)$ can be split into two distinct modular multiplications, it can be realized by two separate sets of $CU$ gates executed on the work register

$$C_m U_{b^{-1}} C_{2m} U_a = \left( \prod_{i=0}^{m-1} C U_{b^{-1}}{}^{2^i} \right) \left( \prod_{i=0}^{2m-1} C U_a{}^{2^i} \right). \quad (19)$$

Therefore, we can also implement the variant iteratively using only a single data qubit: First, we completely process the first multiplication originally controlled by data register $|r_{d_1}\rangle$ realizing $U_a$ and its QFT on a single data qubit. Then, we continue iteratively with the second multiplication originally controlled by the data register $|r_{d_2}\rangle$, its $CU$ operations of the form $U_{b^{-1}}$, and its QFT using the same single data qubit. The downside of the approach stays the same: as all steps are executed sequentially, it introduces idle time between all $CU$ operations and consequentially is the design with the longest runtime.

The alternating design of the standard variant can also easily be transferred to the discrete logarithm variant. For the first multiplication, we use two data qubits alternating between the $CU$ gates of $U_a$ and the phase processing of the QFT of $|r_{d_1}\rangle$. Once the first multiplication is finished, we continue with the second multiplication on the same two data qubits processing the $CU$ gates of $U_{b^{-1}}$ and the phase processing of the QFT of $|r_{d_2}\rangle$.

In the standard variant of the algorithm, the order of executions for the $CU$ operations is forced by the dependencies of the QFT. The $CU^{2^i}$ has to be executed before $CU^{2^{i+1}}$ as its phase correction $P_{i+1}$ depends on the classical information from the measurement outcome following $P_i$. While this holds for the individual multiplications $CU^{2^i}$ inside $U_a$ and $U_{b^{-1}}$, the order of the entire multiplications can be exchanged as they commute. Since the computation on the two control data registers ends with two independent QFT$^\dagger$, no dependence due to phase processing is introduced between $U_a$ and $U_{b^{-1}}$.

While we have to keep the order of the set of $CU_a{}^{2^i}$ as well as the set of $CU_{b^{-1}}{}^{2^i}$ fixed, we do not have to fully process one set before the other, i.e., we can execute a subset of $CU_{b^{-1}}{}^{2^i}$ between any two $CU_a{}^{2^i}$ and $CU_a{}^{2^{i+1}}$ operations and vice versa. Based on this insight, we propose a different design with two data qubits. Again, we alternate between the two data qubits such that a $CU$ operation controlled by one of the two data qubits is always followed on the work register by a $CU$ operation controlled by the other data qubit.

We now propose an adjusted scheduling strategy: the first data qubit performs the first $m$ $CU$ operations of $U_a$ and the second data qubit handles all the $m$ $CU$ operations of $U_{b^{-1}}$. Since $U_a$ has $2m$ $CU$ operations, we execute the last $m$ $CU$ operations of $U_a$ with the usual alternating approach [see

**FIGURE 15.** **Three-cyclic design for the discrete logarithm variant of Shor's algorithm reducing the idle time experienced in the work register by evenly distributing the *CU* operations of the two multiplications $U_a$ and $U_{b^{-1}}$, using three data qubits.**

(20)]. We name this approach the *double-iterative* design as the two multiplications are both iteratively processed on the individual data qubits

$$C_m U_{b^{-1}} C_{2m} U_a = \left(\prod_{i=m}^{2m-1} CU_a{}^{2^i}\right)\left(\prod_{i=0}^{m-1} CU_{b^{-1}}{}^{2^i} CU_a{}^{2^i}\right). \tag{20}$$

This order ensures that, for the first $m$ pairs of $CU$ operations, a phase processing $P_i$ does not depend on measurement information of $P_{i-1}$, but only on the $P_{i-2}$ phase processing. This is the last phase processing that took place on the same data qubit.

Finally, to address the different counts of $CU$ gates for $U_a$ and $U_{b^{-1}}$, we propose a design with three data qubits for the discrete logarithm variant, extending the double-iterative approach. Since there are $2m$ $CU$ operations for $U_a$, but only $m$ for $U_{b^{-1}}$, we reserve two data qubits for $U_a$ and a single data qubit for $U_{b^{-1}}$. We then cycle through the three data qubits, where the first two data qubits control the next $CU$ operations of $U_a$ and the last data qubit controls the $CU$ operations of $U_{b^{-1}}$, as described in the following equation:

$$C_m U_{b^{-1}} C_{2m} U_a = \left(\prod_{i=0}^{m-1} CU_{b^{-1}}{}^{2^i} CU_a{}^{2^{2i+1}} CU_a{}^{2^{2i}}\right). \tag{21}$$

This *three-cyclic* design, as shown in Fig. 15, can further mitigate idle time at the cost of an additional qubit.

### C. DISTRIBUTED SHOR

Several early works have explored distributed implementations of Shor's algorithm. A notable foundational study [69] introduces a distributed approach using VBE [7] adders within modular exponentiation circuits. It assumes multiple nodes (QPUs), each with $n + c$ qubits (where $c$ accounts for ancilla and communication qubits), dividing the $m = 2n$-qubit data register into two blocks across the nodes and splitting the work register into five parts to fit the VBE structure. This work establishes an upper bound of $O(n^2)$ on communication overhead, with gate complexity $O(n^3)$ and space overhead $7n + 1$ for factoring an $n$-bit number.

A comparative study [70] evaluates various adder implementations under different distributed settings, including qubit capacities, network topologies, and I/O capabilities. It contrasts EJPP and teledata protocols, demonstrating that the teledata method outperforms EJPP by reducing communication overhead and enabling better parallelism, especially favoring carry-ripple adders for large-scale problems.

More recently, Neumann et al. [71] examines the effects of imperfect ebits in distributed QPE using a QFT with embedding techniques. This study highlights that aggregating multiple nonlocal operations over a single EJPP channel can reduce communication overhead from $O(n^2)$ to $O(n)$ and improve fidelity in implementing an $n$-qubit distributed QFT.

Meanwhile, Gidney and Ekerå [5] propose circuit-level optimizations for monolithic implementations of Shor's algorithm targeting large integers, focusing on modular exponentiation. Techniques such as coset representations, windowed arithmetic, oblivious carry runways, and lookup-based modular addition significantly reduce time and qubit requirements. While mainly monolithic, this work acknowledges distributed implementations as promising future enhancements.

On the other hand, general-purpose distributed compilation frameworks [19], [20], [72], [73] do not explicitly exploit Shor's algorithm structure, like the distinct roles of data and work registers, which may limit optimization potential. For instance, Ferrari et al. [19] address the challenges in distributed quantum compiling—primarily inter-QPU communication overhead and suboptimal solutions—by proposing a general, efficient, and effective quantum compiler, validated through performance evaluations.

Building on this, Ferrari et al. [20] present a modular compiler optimized for distributed environments, balancing communication costs and device constraints via cost-based scheduling between EJPP and teledata protocols. Qubit assignment uses METIS-based graph partitioning with local refinements. Evaluations on circuits like VQE and QFT across diverse network topologies reveal that teledata's impact on EPR consumption depends on circuit structure.

The work in [72] extends the EJPP protocol to a generalized entanglement-assisted packing scheme, minimizing entanglement and auxiliary qubits by leveraging packing and conflict graphs. Applied to unitary coupled-cluster (UCC) circuits, it achieves substantial entanglement savings and approaches theoretical lower bounds.

Similarly, Andres-Martinez et al. [73] adapt embedding-based gate packing to heterogeneous multipartite networks, reducing entanglement costs through Steiner-tree-based entanglement sharing. Tested on multiple network types, this method significantly lowers EPR usage, particularly in quantum chemistry circuits like UCC.

In contrast to these prior efforts, which either focus on low-level circuit partitions or general compilation, our proposed distributed designs operate at a higher abstraction level and remain agnostic to specific adder choices. Based on this criterion, we avoid partitioning the work register. Instead, we place the entire data register on one QPU and the work register on another (see Fig. 16). This approach maintains architectural flexibility and is essential for potential integration with finer-grained partitioning of the work register at lower layers of the design stack. Any further splitting of the work register will be adder-dependent, and each part of these split modular arithmetic operations will require distributed control information from the data register. The cost of such a low-level splitting is then the cost of teleportation and EJPP protocols necessary to distribute the arithmetic circuit (e.g., two teleportations to split a VBE adder or six teleportations to split a CDKM adder [70]) and additionally the cost of data register distribution. From this perspective, our investigated splitting is the most fundamental and ebit-efficient splitting to distribute Shor's algorithm.

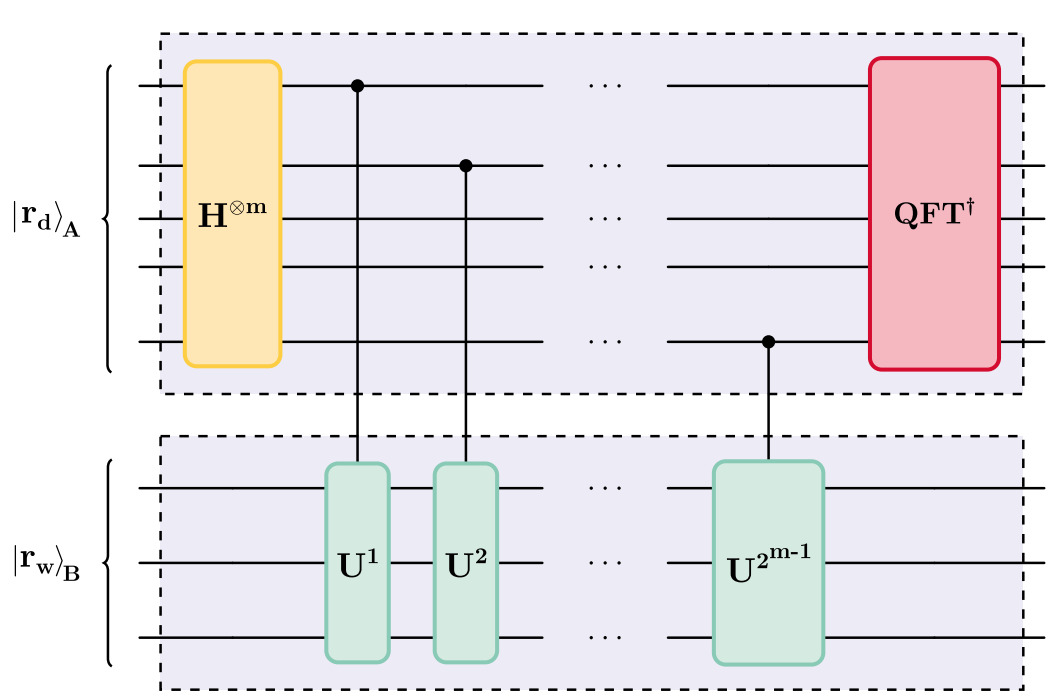

**FIGURE 16.** Distributed design of Shor's algorithm: the data register $|r_d\rangle$ is allocated to QPU $A$, while the work register $|r_w\rangle$ is allocated to QPU $B$. The $CU^{2^i}$ operations must be performed remotely between the two QPUs.

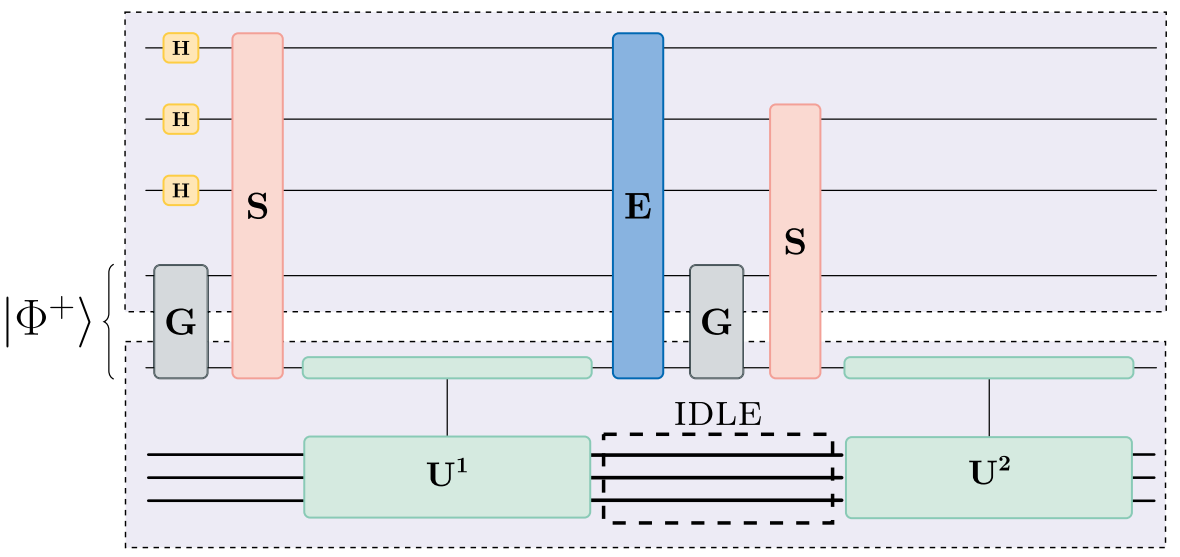

**FIGURE 17.** Regular distributed design of Shor's algorithm results in idle time between two $CU^{2^i}$ operations due to the usage of a single ebit channel. The idle period occurs due to the end process ($E$) of the previous $CU^{2^i}$ operation, as well as the ebit generation ($G$) and start process ($S$) for the next $CU^{2^{i+1}}$ operation.

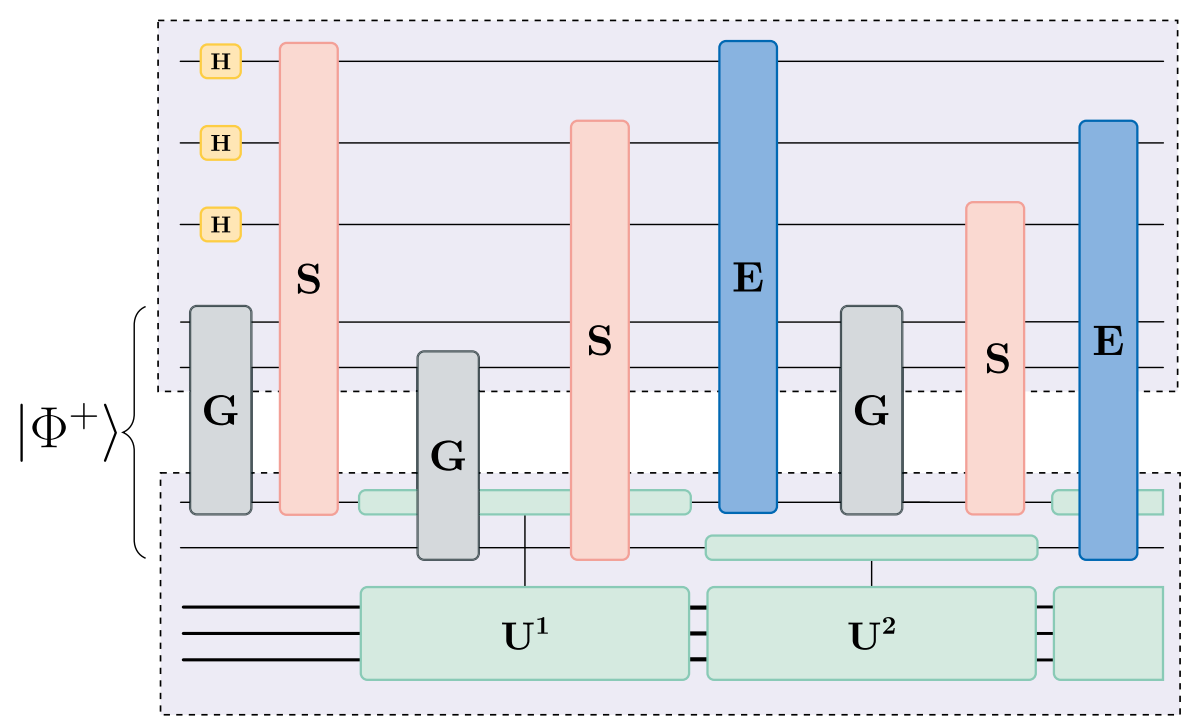

**FIGURE 18.** Regular distributed design of Shor's algorithm utilizing two ebit channels to eliminate idle time in the work register caused by ebit generation ($G$), start process ($S$), and end process ($E$). The idle period reduction is achieved by alternating the two ebit channels, enabling the scheduling of distribution processes to occur simultaneously to $CU^{2^i}$ operations.

### D. DISTRIBUTED PARALLELIZATION

In Shor's algorithm, since the QFT$^\dagger$ operates entirely on the data register, the primary task for distribution involves modular exponentiation, which affects both the data and work registers. According to the $C_mU$ decomposition from (7), the $CU^{2^i}$ gates, each controlled by a single qubit on the data register, must be executed across separate QPUs. Teleporting the control qubits to the work QPU and then back to the data QPU would be costly, requiring two ebits per $CU^{2^i}$ gate, amounting to a total of $2m$ ebits. Instead, we use the EJPP protocol to perform each $CU^{2^i}$ operation remotely. This process consists of three steps: 1) applying a starting process $S$, which involves the control data qubit for the $CU^{2^i}$ gate on QPU $A$ and a shared ebit between QPUs $A$ and $B$; 2) executing the $CU^{2^i}$ gate locally on the work register located on QPU $B$ using the shared ebit; and 3) completing the remote operation with an ending process $E$, which uses the control qubit from QPU $A$ and the shared ebit. The entire process can be represented as

$$CU\,|r_d\rangle\,|r_w\rangle \rightarrow E(CU)S\,|r_d\rangle\,|\phi^+\rangle\,|r_w\rangle\,. \tag{22}$$

With embedding, only one ebit is needed for each $CU^{2^i}$ operation, requiring a total of $m$ ebits, even when the $CU$ gates are decomposed into native gates. The $S$ and $E$ operations involve only a few basic gates and introduce minimal delay. However, generating an ebit ($G$) between two QPUs can be time consuming, depending on the hardware and distance. Since the distribution setup requires $m$ ebits, multiple ebit generation cycles will be necessary during the distributed execution of the algorithm.

Furthermore, if only a single ebit channel is available between the two QPUs, the previous EJPP protocol must be completed with an ending process before a new ebit is generated over the ebit channel, and the starting process begins the next EJPP protocol. Similar to the iterative monolithic design, the work register remains idle due to the inclusion of processes $G$, $E$, and $S$ in the critical path. Fig. 17 illustrates the regular distributed design of Shor's algorithm and highlights the instances of idle time that occur during the scheduling of $CU^{2^i}$ operations on the work register when only a single ebit channel is used.

When multiple ebit channels are available, distribution tasks can be parallelized, as illustrated in Fig. 18 for the regular design. While one $CU^{2^i}$ operation is in progress, the ebit for the next $CU^{2^{i+1}}$ can be generated, and the starting process for the next operation can be executed. Since the next $CU^{2^{i+1}}$ works on a different data qubit, it can proceed while the current $CU^{2^i}$ is running. Furthermore, the next $CU^{2^{i+1}}$ does not need to wait for the ending process of the previous one, as it can be processed in parallel. By alternating between two ebit channels, similar to the alternating design for the monolithic setup, the idle time in the work register can be reduced.

The benefit of using more ebit channels depends on both the delay of the $CU^{2^i}$ operations and the time required for ebit generation. If the ebit generation process is too slow, it cannot run fully in parallel with a $CU^{2^i}$ gate and will delay the next $CU^{2^{i+1}}$ operation. With two ebit channels, during the time of one $CU^{2^i}$ gate, the start process ($S$), ebit generation ($G$), and end process ($E$) must all be completed to avoid idle time. With three ebit channels, the next ebit must be ready after the execution of two gates, $CU^{2^i}$ and $CU^{2^{i+1}}$. Likewise, with four ebit channels, the next ebit must be ready after three gates, $CU^{2^i}$, $CU^{2^{i+1}}$, and $CU^{2^{i+2}}$, and so on. This sets the condition for zero idle time with $k$ ebit channels, meaning that a $CU^{2^{i+k}}$ operation will not be delayed if the full distribution block ($G$, $S$, $E$) has been executed during all the previous $k-1$ gates, $CU^{2^i}$, $CU^{2^{i+1}}$,..., $CU^{2^{i+k-1}}$:

$$\sum_{j=0}^{k-1} t_{CU^{2^{i+j}}} \geq t(ESG) \quad \forall i. \tag{23}$$

For large $m$, a looser bound can be derived by using the average $CU^{2^i}$ gate delay $\bar{t}_{CU}$, which approximates (23) for large $m$

$$(k-1)\bar{t}_{CU} \geq t(ESG). \tag{24}$$

We investigate the amount of idle time from distribution that can be mitigated by analyzing how the delay of distribution processes contributes to the overall circuit delay of (15)

$$t_C = t_H + \sum_{i=0}^{m-1} t_{CU^{2^i}} + \delta_P + \delta_D. \tag{25}$$

The initial Hadamard and the $CU^{2^i}$ operations still have to be executed sequentially, so their delays will always contribute to the overall circuit delay and there can be delay from phase processing. The additional delay based on distribution is denoted as $\delta_D$ and depends on the amount and speed of EJPP channels used, as well as the duration of $CU^{2^i}$ operations. We split $\delta_D$ into mitigatable delay $\delta_D^M$ and unavoidable delay $\delta_D^{\neg M}$ due to distribution

$$\delta_D = \delta_D^M + \delta_D^{\neg M}. \tag{26}$$

The ebit generation and start process for the first $CU^{2^i}$ gate will inevitably contribute to the circuit delay, as they cannot be parallelized. Similarly, the final termination process cannot be executed concurrently with the last $CU^{2^{m-1}}$ gate. Combined, the unavoidable delay is equal to the delay of one full distribution block ($G$, $S$, $E$)

$$\delta_D^{\neg M} = t(SG) + t_E = t(ESG). \tag{27}$$

For the other distribution operations, if the other operations are long enough and the critical path goes through the $CU^{2^i}$ gates or phase processing, no delay of the distribution operation processing contributes to the circuit delay. However, if the $CU^{2^i}$ and phase processing operations are instant or not running in parallel, all distribution delay contributes. Therefore, the mitigatable delay from distribution processes is bound by

$$\begin{aligned} 0 \leq \delta_D^M &\leq t_E + \sum_{i=1}^{m-2} t(ESG) + t(SG) \\ &= (m-1)t(ESG). \end{aligned} \tag{28}$$

In principle, these bounds apply to all setups with multiple ebit channels and all three distributed designs (i.e., regular, iterative, and alternating) in the standard variant, as well as all designs of the discrete logarithm variant. However, for the iterative design, it is worth noting that the start and end processes cannot be parallelized because all operations ($S$, $E$, and $CU^{2^i}$) act on the same single data qubit. Note that only in the distributed iterative design (i.e., characterized by sequential execution on a single-qubit data register, as shown in Fig. 12) with a single ebit channel, the delays associated with phase processing (as defined by (8), followed by a Hadamard and measurement) and with the EJPP protocol stages ($G$, $S$, and $E$, as shown in Fig. 4) contribute independently to the total circuit delay. However, if multiple data qubits are used, phase processing can run in parallel with ebit generation and CU gates on the work register, reducing the overall idle time. Similarly, when multiple ebit channels are used, the distribution process can be parallelized, and the delay contributions $\delta_P$ and $\delta_D$ are highly dependent on one another. In these cases, the total idle time depends on which operation becomes the bottleneck in the computation, determining the critical path.

## IV. METHOD

To evaluate the impact of our proposed design approaches on circuit delay, we selected modular exponentiation circuits from available resources and implemented them using the Qiskit framework [74]. Since the performance of the designed circuits depends on the latency of primitive operations supported by the targeted hardware, we constructed both monolithic and DQC models in accordance with existing literature and available quantum systems [23], [29], [30], [31], [48], [49], [51], [63], [75], [76], [77], [78], [79], [80], [81], [82], [83], [84], [85], [86], [87], [88], [89], [90], [91], [92], [93], [94], [95], [96], [97].

More specifically, and starting with `Qrisp` [23], we used this library to automate the construction of the $CU$ operations in Shor's algorithm. This allowed us to scale our analysis up to $n = 64$ bits, which is the current version's limit. The automated compilation process through `Qrisp` allows us to conduct an analysis across multiple values of $n$, contrary to [29] and [30].

Finally, in order to conduct a hardware-faithful analysis, we base our parameter values on the available hardware as well as the available hardware literature. We start by considering the parameter values based on the commercially available quantum computers through cloud access [75], [76]. In these cases, the necessary hardware parameters for superconducting and ion-trap systems, such as gate, measurement, and reset times, are readily available. For the neutral atom case, we instead compiled a list of the necessary parameter values from [48], [49], [51], [78], [79], [80], [81], [82], [83], [84], and [85], which is summarized later on in Table IV. This covers the necessary hardware parameters for the monolithic case. For the DQC case, we used the large body of existing work to model the ebit generation time [63], [86], [87], [88], [89], [90], [91], [92], [93], [94], [95], [96], as detailed in Section IV-B.

**TABLE 2.** Average Delays for Single-Qubit Gates ($t_{q_1}$), Two-Qubit Gates ($t_{q_2}$), and Reset ($t_{\text{reset}}$) and Measurement ($t_{\text{measure}}$) Times for Superconducting QPUs From IBM [75]

| Superconducting Backend Properties | | | | |
|---|---|---|---|---|
| Backend | $t_{q1}$ | $t_{q2}$ | $t_{\text{measure}}$ | $t_{\text{reset}}$ |
| Eagle sherbrooke | 57 ns | 533 ns | 1216 ns | 1276 ns |
| Heron r1 torino | 32 ns | 68 ns | 1560 ns | 1708 ns |
| Heron r2 fez | 24 ns | 84 ns | 1560 ns | 1584 ns |
| Heron r2 marrakesh | 36 ns | 68 ns | 2100 ns | 2236 ns |

**TABLE 3.** Average Delays for Single-Qubit Gates ($t_{q_1}$), Two-Qubit Gates ($t_{q_2}$), as Well as Reset ($t_{\text{reset}}$) and Measurement ($t_{\text{measure}}$) Times for Ion-Trap QPUs From IonQ [76]

| Ion-Trap Backend Properties | | | | |
|---|---|---|---|---|
| Backend | $t_{q1}$ | $t_{q2}$ | $t_{\text{measure}}$ | $t_{\text{reset}}$ |
| IonQ Aria 1 | 135 $\mu s$ | 600 $\mu s$ | 300 $\mu s$ | 20 $\mu s$ |
| IonQ Aria 2 | 135 $\mu s$ | 600 $\mu s$ | 50 $\mu s$ | 15 $\mu s$ |
| IonQ Forte | 130 $\mu s$ | 970 $\mu s$ | 150 $\mu s$ | 50 $\mu s$ |

### *A. QPU MODELING*

For both monolithic and distributed setups, we adopted a generic basis gate set $\{I, X, H, P(\theta), CX\}$, requiring transpilation only for gates with multiple controls and SWAP gates. In addition, we did not impose a specific qubit topology, eliminating the need for extra SWAP gates to enable two-qubit interactions. The discussed parallelization approaches are compatible with any quantum hardware platform. Three widely used platforms were selected for the experiments—superconducting, ion-trap, and neutral-atom-based QPUs—where ebit generation has been experimentally demonstrated and dynamic circuits are supported. For superconducting QPUs, we used IBM Eagle and IBM Heron characteristics as benchmarks for gate execution times, measurements, and resets, leveraging their accessibility through the IBM Quantum Platform [75]. To map gate times to our chosen generic basis gate set, we computed the average single-qubit gate time ($t_{q_1}$) and two-qubit gate time ($t_{q_2}$). A detailed list of extracted backend properties is provided in Table 2.

For ion-trap QPUs, we similarly extracted execution times for the IonQ Aria and IonQ Forte systems from the IonQ cloud [76], as reported in Table 3.

Regarding neutral atom QPUs, while QuEra systems [77] are available via cloud access, they are restricted to analog quantum computing and do not support quantum circuits.

**TABLE 4.** Reported Average Delays for Single-Qubit Gates ($t_{q_1}$), Two-Qubit Gates ($t_{q_2}$), and Reset ($t_{\text{reset}}$) and Measurement ($t_{\text{measure}}$) Times for Neutral Atom Quantum Computers in the Literature

| Neutral Atom Backend Properties | | | | |
|---|---|---|---|---|
| Backend | $t_{q1}$ | $t_{q2}$ | $t_{\text{measure}}$ | $t_{\text{reset}}$ |
| [78] | 250 ns/4.1 $\mu s$ | 416 ns | 6 ms | N/A |
| [51] | $\sim$2 $\mu s$ | $\sim$400 ns | $\sim$10 ms | N/A |
| [82] | N/A | $\sim$250 ns | N/A | N/A |
| [83] | N/A | $\sim$110 ns | N/A | N/A |
| [84] | N/A | $\sim$6.5 ns | N/A | N/A |
| [49] | N/A | N/A | 4 ms | N/A |

**TABLE 5.** Average Delays for Single-Qubit Gates ($t_{q_1}$), Two-Qubit Gates ($t_{q_2}$), and Reset ($t_{\text{reset}}$) and Measurement ($t_{\text{measure}}$) Times for the Selected Experimental Setups: Ion-Trap (IonQ Forte), Superconducting (IBM Heron), and Neutral Atom Setups Chosen for Experiments

| Backend Properties | | | | |
|---|---|---|---|---|
| Backend | $t_{q1}$ | $t_{q2}$ | $t_{\text{measure}}$ | $t_{\text{reset}}$ |
| IonQ Forte | 130 $\mu s$ | 970 $\mu s$ | 150 $\mu s$ | 50 $\mu s$ |
| IBM Heron | 32 ns | 68 ns | 1560 ns | 1708 ns |
| Neutral Atom | 2 $\mu s$ | 400 ns | 10 ms | 10.002 ms |

Therefore, we based our model (see Table 4) on recent advancements in gate-based quantum computing from the literature [48], [49], [51], [78], [79], [80], [81], [82], [83], [84], [85].

Finally, we picked one representative system from each platform for further analysis (see Table 5): IonQ Forte for ion-trap QPUs, IBM Heron (ibm_torino) for superconducting QPUs, and a neutral atom QPU based on the reports in [51].[2] We argue that it is natural to include an IonQ backend as well as an IBM Heron backend in our final selection. For IBM Heron, this is planned as a DQC architecture. Moreover, IonQ also plans a distributed quantum architecture as part of their short-to-medium-term roadmap, which makes our analysis timely.

### *B. DQC MODELING*

As there is no DQC hardware publicly available yet, we cannot directly extract execution times for distributed operations. Instead, we have to rely on models based on experimental DQC setups. We assume a distribution model where QPUs are interconnected via quantum and classical channels, with ebits facilitating nonlocal gate execution. Our parallelization approach requires only two QPUs and does not assume a specific network architecture. For simplicity, we consider a direct connection between the two systems, eliminating the need for routing through quantum nodes, quantum repeaters, or ES.

Since Qiskit does not yet support a distributed circuit model, we constructed circuits in which qubits are divided into subsets corresponding to the first and second QPUs, along with qubit pairs designated for ebit channels. The ebit generation process is modeled using a Hadamard and a CNOT gate, with modified execution times ($t_{\text{ebit}_H}$, $t_{\text{ebit}_{CX}}$) representing the ebit generation time ($t_{\text{ebit}}$).

[2] The reset duration for the neutral atom QPU model is modeled as the combined time of a measurement followed by a single-qubit gate, as no specific values have been reported.

The ebit generation between distant QPUs has been experimentally demonstrated using photonic quantum communication, coupled with computing platforms based on atoms [86], [87], [88], [89], [90], ion traps [91], [92], [98], diamonds [93], and superconductors [94], [95], [96]. Based on these studies, we calculated a realistic range of ebit generation times ($t_{\text{ebit}}$) for our three chosen hardware platforms, as shown in the following paragraphs.

Our long-range ebit generation model is based on [63], which assumes a neutral atom platform. In this model, we assume a local heralded entanglement generation between a telecom photon and a local qubit for both QPUs. The probability of successful local entanglement generation is given by

$$p = p_{ht}\nu_h\nu_t \tag{29}$$

where $p_{ht}$ is the photon generation probability, $\nu_h$ is the heralding detector efficiency, and $\nu_t$ is the entangling detector efficiency. After both QPUs send their entangled photons, a photonic Bell state measurement (BSM) at a midpoint between them generates the end-to-end ebit. The success probability of this process is

$$p_e = \frac{1}{2}\nu_o p^2 e^{\frac{-d}{L_0}} \tag{30}$$

where $\nu_o$ is the optical BSM efficiency, $d$ is the link length, and $L_0$ is the attenuation length of the optical fiber. The average time for a successful ($T_s$) and failed ($T_f$) entanglement attempt is given by

$$T_s = \tau_p + \max\left\{\tau_h, \tau_t + \frac{d}{c_f} + \tau_o\right\} \tag{31}$$

$$T_f = \tau_p + \max\left\{\tau_h, \tau_t + \frac{d}{c_f} + \tau_o, \tau_c\right\} \tag{32}$$

where $\tau_p$ is the time required to excite the qubit, $\tau_h$ is the herald-cavity time, $\tau_t$ is the telecom-cavity time, $\tau_o$ is the optical BSM time, $c_f$ is the speed of light in the optical fiber, $\frac{d}{c_f}$ accounts for the time it takes for the telecom photons to reach the midpoint and for the classical acknowledgment being sent back from the BSM, and $\tau_c$ represents the additional time required for a local qubit reset in case of a failed entanglement attempt.

Overall, the expected entanglement generation time is then calculated as

$$T = \frac{p_e T_s + (1 - p_e)T_f}{p_e}. \tag{33}$$

For neutral atom systems, we use parameter values commonly found in the literature [86], [87]

$$p_{ht} = 0.53, \quad \nu_h, \nu_t = 0.8, \quad \nu_o = 0.39,$$
$$L_0 = 22 \text{ km}, \quad \tau_o = \tau_a = \tau_t = 10\,\mu\text{s}, \quad c_f = 2\times\ 10^8 \text{ m/s},$$
$$\tau_p = 5.9\,\mu\text{s}, \quad \tau_h = 20\,\mu\text{s}, \quad \tau_d = 100\,\mu\text{s}.$$

For distances ranging from 1 to 50 km, these parameters yield ebit generation times between 5 and 115 ms.[3] A similar heralding approach has been demonstrated for superconducting systems [95], with reported ebit generation times between 10 and 1000 $\mu$s. For ion-trap QPUs, ebit generation times of approximately 5.5 ms have been observed over short distances [92], while for longer distances (up to 230 m), the reported range is 2–17 s [91].

These values represent optimistic estimates, as our model does not account for additional factors such as ES, quantum repeaters, or entanglement purification, all of which would reduce the success probability of end-to-end ebit generation and increase the overall generation time.

### C. MODULAR EXPONENTIATION CIRCUITS

Only a limited set of synthesized circuits for Shor's algorithm are currently available. In contrast to conventional designs that build modular multiplication using general-purpose adder circuits, previous works [29], [30], [31] have introduced highly optimized circuits specifically tailored for factoring small values of $N$. These designs prioritize reduced circuit width and depth, enabling early experimental demonstrations on NISQ hardware. In addition, MQT Bench [97] provides fixed Shor circuits for $N = 9, 15, 821$ as benchmarks for design evaluations.

For our experiments, we utilized the QRISP framework [23], which offers a high-level interface for generating modular multiplication circuits for arbitrary $N$, supporting various arithmetic adders [8], [11], [12], [99]. We employed the default Fourier adder [12] in our implementations to ensure that the evaluation remained focused on assessing the effectiveness of the proposed design relative to traditional and iterative variants. As the proposed methods are agnostic to specific arithmetic constructions, they can seamlessly integrate with any of the aforementioned adders.

For the optimized circuits, we implemented simplified $CU$ operations by following the design guidelines provided in the relevant literature [29], [30], [31]. In the case of QRISP, the underlying factorization function was used to extract $CU$ operations, which were then incorporated into the Qiskit [74] environment. For the discrete logarithm variant, $CU_{b^{-1}}$ operations were constructed similarly. Since $b^{-1} \mod N$ and its powers can be precomputed classically, they were supplied to QRISP to synthesize the corresponding modular multiplication circuits.

Finally, although multiple coprime values of $a$ are available for any given $N$ (e.g., for $N = 15$, valid choices include $a = 2, 7, 8, 13$) in the modular exponentiation step [see (1)], the specific choice of $a$ can influence both circuit width and runtime. To ensure consistency and maintain a manageable scope for evaluation, we fixed $a = 2$ across all experiments.

[3] We assume no coherence time restrictions, as in [63], which could otherwise further constrain the capacity of an ebit channel.

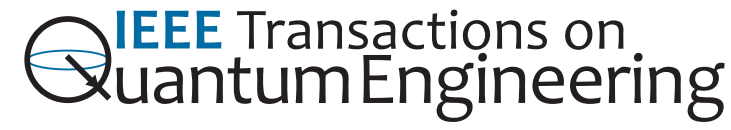

### *D. DELAY CALCULATION*

To estimate circuit delay in practice, a WDAG representation [see Fig. 6(c)] is often employed. This model introduces dummy source (Sc) and sink (Sk) vertices at the beginning and end of the graph, respectively. The circuit delay is then computed as the length of the longest path from *Sc* to *Sk*.

In contrast to the WDAG definition, where instruction weights are assigned to vertices, standard weighted path analysis typically places weights on edges. Without loss of generality, however, vertex weights can be shifted to their incoming edges $(., v)$, since any path traversing vertex $v$ must pass through at least one of its incoming edges. As the graph is acyclic, each path will use at most one such edge. To model initial and terminal instructions accurately, the dummy source and sink nodes are used to distribute weights over incoming and outgoing edges, respectively.[4]

For DAG-based circuit representations, the longest path can be efficiently computed using topological sorting. When quantum circuits are generated programmatically (e.g., in Qiskit) operations are typically appended in order, inherently creating a topological ordering of the DAG. This allows a direct application of the longest path algorithm. The space complexity of this computation is $O(V)$, where $V$ is the number of vertices. If only the critical path length is needed (not the full path), it suffices to store just the distances from the source node.

In quantum circuits, this process can be further optimized by tracking delay information only per qubit and classical bit, rather than per operation. Since each instruction acts on a limited number of bits, the number of relevant incoming edges is bounded by the number of available bits. By associating each edge with the specific qubits or bits it affects, memory usage can be significantly reduced. This optimization is implemented in the space-efficient critical path algorithm shown in Algorithm 1.

In our experiments, we applied various weight mappings to reflect realistic gate durations. Special operations, such as reset, measure, and ebit-related instructions ($\text{ebit}_\text{h}$, $\text{ebit}_\text{CX}$), were assigned distinct weights, while standard single- and two-qubit gates were modeled using average durations $t_{q_1}$ and $t_{q_2}$. Since ebit generation is abstracted via a Hadamard and CNOT gate (rather than a full physical implementation), we set $\text{ebit}_\text{h} = 0$ and allocate the entire execution time to $\text{ebit}_\text{CX}$, as outlined in Section IV-B. This ensures a balanced execution time distribution across the two qubits involved in the ebit channel.

To evaluate execution time, we considered three approaches: circuit moments, WDAG-based delay using NetworkX [100], and the optimized method in Algorithm 1. While circuit moments and WDAG representations offer additional features beneficial to circuit compilation workflows, Algorithm 1 was the most computationally efficient.

[4]To accurately represent the weights of the initial instructions, we utilize the dummy source node ($s_k$). Similarly, the outgoing edges $(v, .)$ can be used to model the weights of the instructions together with the dummy sink node ($s_k$).

**Algorithm 1:** Circuit Delay.

**Input:**
  Circuit as topologically sorted list:
  $U = [U_1, \ldots, U_{N_U}]$,
  List of circuit qubits and clbits: $B$
**Result:** Circuit delay $t_C$
$t[b] = 0, \forall b \in B$
**for** $i = 1, \ldots, N_U$ **do**
  $B_i \leftarrow U[i]$.bits
  $t_i \leftarrow U[i]$.delay
  $t_{\text{max}} \leftarrow \max(\{t[b], \forall b \in B_i\})$
  **foreach** $b \in B_i$ **do**
    $t[b] = t_{\text{max}} + t_i$
  **end**
**end**
$t_C \leftarrow \max(\{t[b], \forall b \in B\})$
**return** $t_C$

As our goal was solely to compute critical path lengths without leveraging these extra features, we adopted Algorithm 1 as the default method for our delay evaluation.

## V. RESULTS

We evaluated our proposed design approach across three distinct hardware platforms: neutral atoms, superconducting qubits, and ion traps. All code with which the circuits were built and the experiments were conducted is available online.[5] The analysis included three circuit sets: optimized circuits [29], [30] for the standard variant of Shor's algorithm and QRISP-generated circuits [23], with $N$ sizes of up to $n = 64$ bits, which were used for both the standard and discrete logarithm variant, as specified in Section IV-C. To the best of our knowledge, we are not aware of any other methods than `Qrisp` to automatically generate the necessary circuits for our analysis. For the evaluation, we utilized the QPU parameters of ibm_torino heron QPU for superconducting qubits, the ionq_forte QPU for ion traps, and neutral atom QPU parameters reported in [51], as discussed in Section IV-A.

Key metrics assessed for each circuit included work register size, gate counts (single- and two-qubit gates), and the average delay for executing controlled unitary (CU) operations on the selected platforms. The results are summarized in Fig. 19. The effectiveness of the proposed approach was evaluated in both monolithic and distributed setups.

In the monolithic environment, experimental results demonstrated that for all circuits, our proposed designs: 1) can achieve the same circuit depth as the corresponding regular design across all architectures and 2) require only one or two additional qubits compared to the iterative design. We highlight differences in delay performance based on qubit technologies and analyze delay scaling behavior for varying sizes of $N$ to be factored. In the distributed environment, experiments revealed that the depth of the proposed circuits was substantially influenced by: 1) the availability of ebit channels and 2) the ebit generation time, with higher ebit generation times requiring multiple ebit channels to maintain performance. Our results show the translation of monolithic behavior to distributed setups and how efficient utilization of distribution hardware can be guided by our proposed bounds. Detailed experimental results are presented as follows.

[5]https://github.com/SchmidtMoritz/DistributedShor

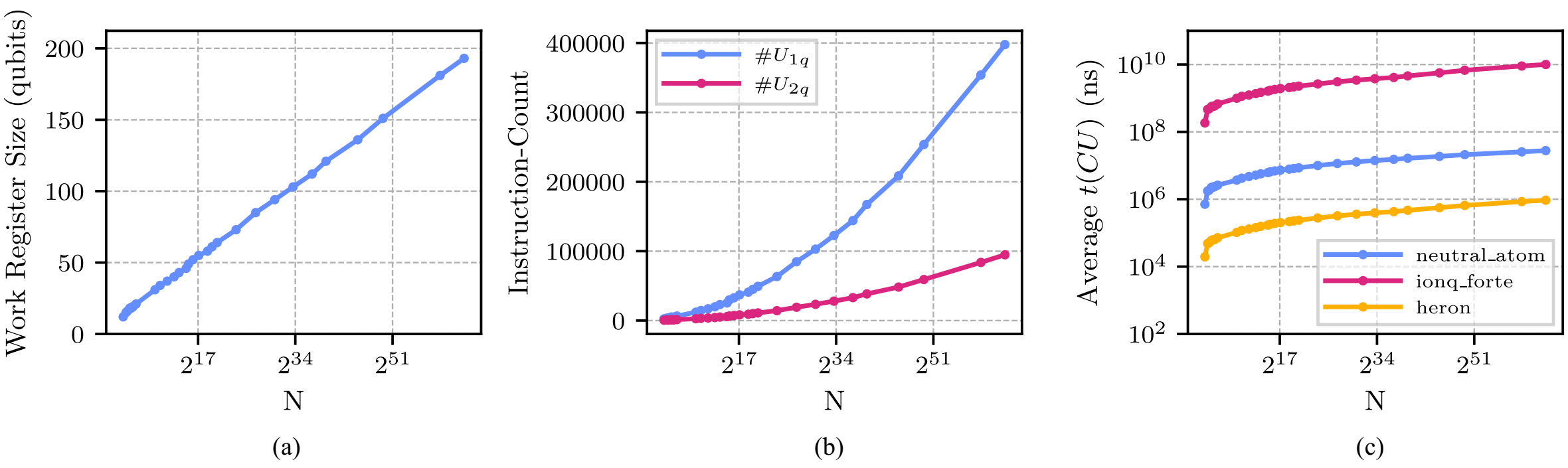

**FIGURE 19.** (a) Size of the work register, (b) single- and two-qubit gate counts, and (c) average delay for the neutral atom, superconducting, and ion-trap weights for the CU gates of the standard variant experiments. The arithmetic circuits were generated using QRISP [23].

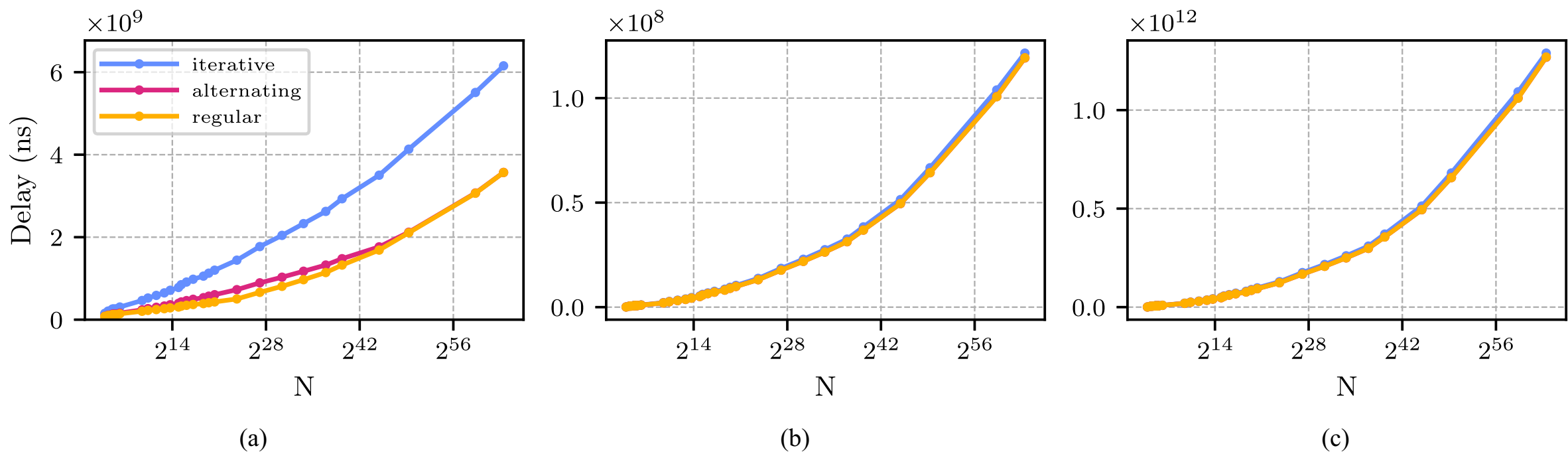

**FIGURE 20.** Monolithic delay scaling of the iterative, alternating, regular approach for (a) neutral atom, (b) IBM superconducting, and (c) IonQ ion-trap weights.

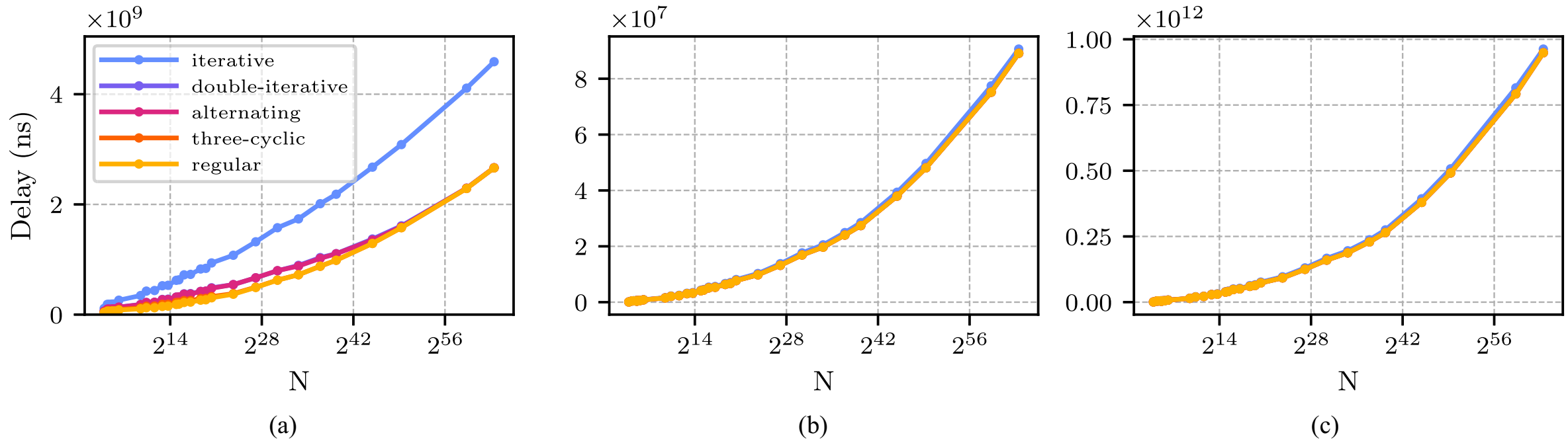

**FIGURE 21.** Monolithic delay scaling of the iterative, double-iterative, alternating, three-cyclic, and regular approach for the discrete logarithm variant of Shor's algorithm and (a) neutral atom, (b) IBM superconducting, and (c) IonQ ion-trap weights.

### *A. ANALYSIS IN THE MONOLITHIC SETUP*

Initially, in the monolithic environment, we constructed iterative, regular, and proposed alternating circuits for the standard variant and additionally double-iterative and three-cyclic circuits for the discrete logarithm variant for all chosen values of $N$ and calculated their delay across architectures, as shown in Figs. 20 and 21. For superconducting and ion-trap QPUs, the delay of $CU$ operations quickly outpaced the delays of phase processing and qubit resets, even for small $N$. Consequently, the mitigable idle time formed only a negligible part of the overall circuit delay, resulting in roughly equal performance across all monolithic designs within these architectures.

In contrast, for neutral atoms, even at the largest tested $N$ (i.e., $n = 64$), a substantial delay difference persisted between iterative and regular/our proposed designs due to (14) being violated. This aligns with expectations, as measurement and reset times for neutral atoms remain orders of magnitude slower than gate operations. Conversely, in superconducting QPUs, measurements and resets are only slightly slower, and in IonQ systems, they are even faster than single- and two-qubit operations. While the iterative design is the most qubit-efficient design, its delay is substantially larger compared to the other approaches. The alternating, double-iterative, and three-cyclic design achieved delays equal to or close to the regular design but required only one additional qubit compared to the iterative approach.

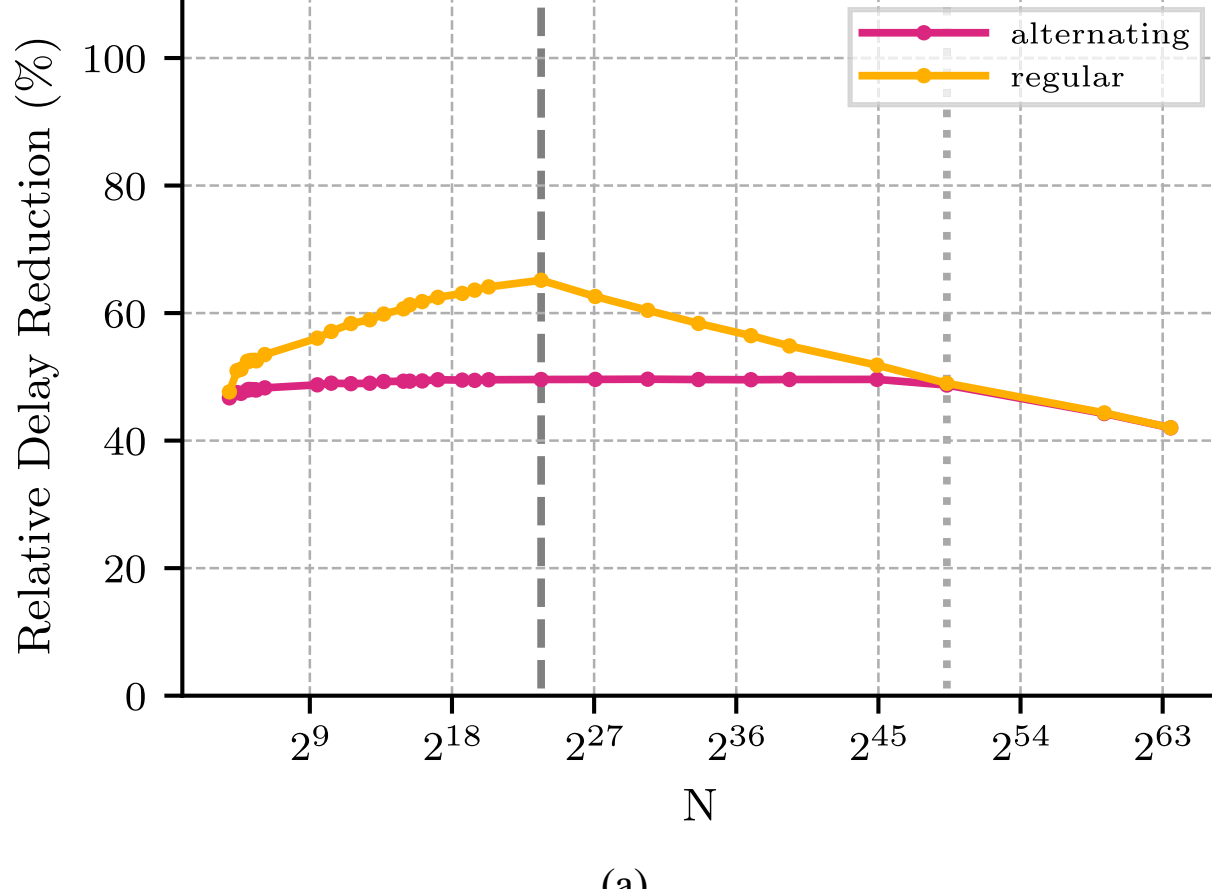

(a)

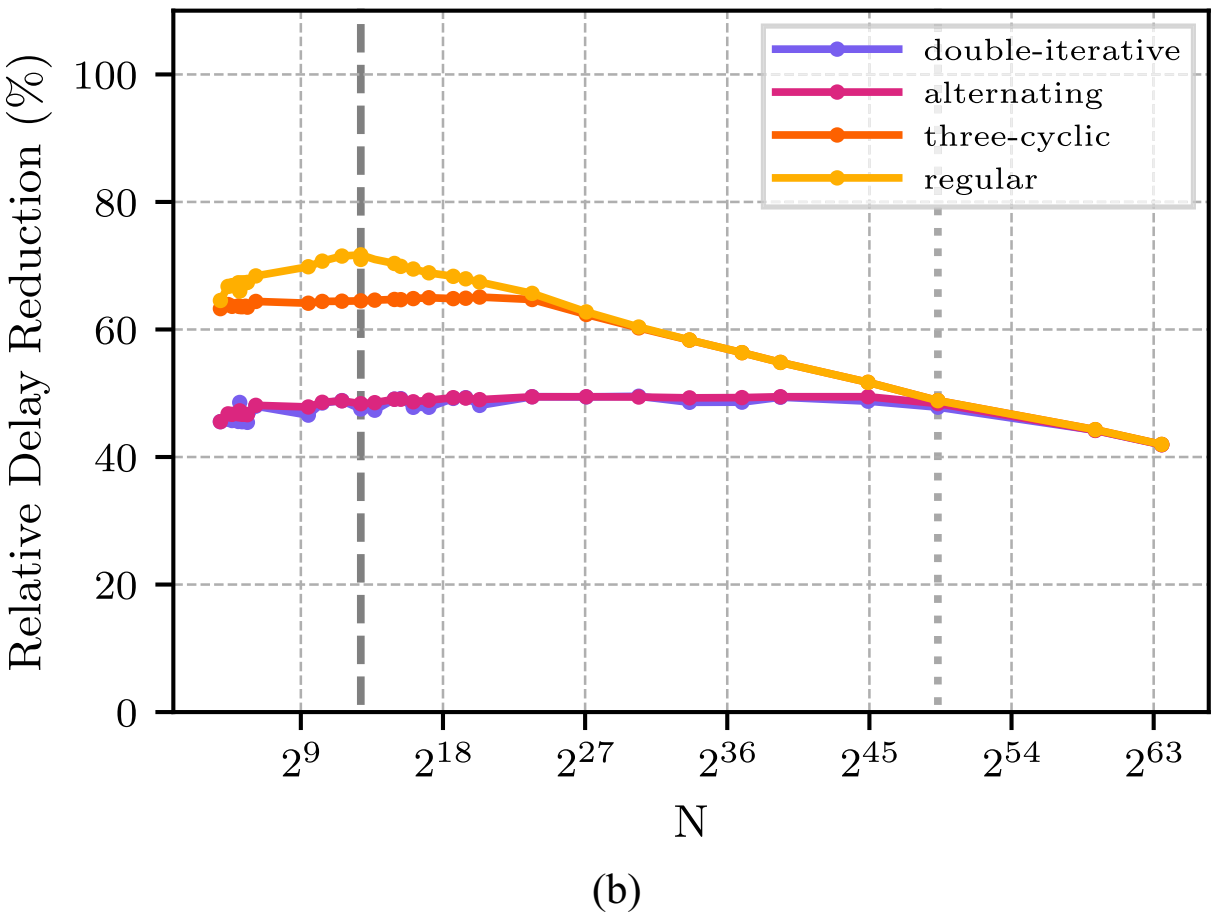

(b)

**FIGURE 22.** **Relative delay reduction of the alternating and regular design compared to the delay of the iterative design for the standard (a) and additionally the double-iterative and three-cyclic for the discrete logarithm variant (b) using the neutral atom QPU weighting.**

To further explore the differences between the regular and our proposed designs in neutral atoms, we computed the relative delay reduction compared to the iterative approach (see Fig. 22). For the standard variant, the alternating design consistently reduced delay by 50% up to $n = 50$. The regular design also began with a 50% reduction for small $N$ but peaked at $n = 25$ before converging with the alternating design's performance at $n = 50$. Beyond $n = 50$, both designs exhibited equal and decreasing delay reductions.

For the discrete logarithm variant, the alternating and double-iterative designs, which both use two data qubits, are performing close to even and are already largely improving over the iterative design.

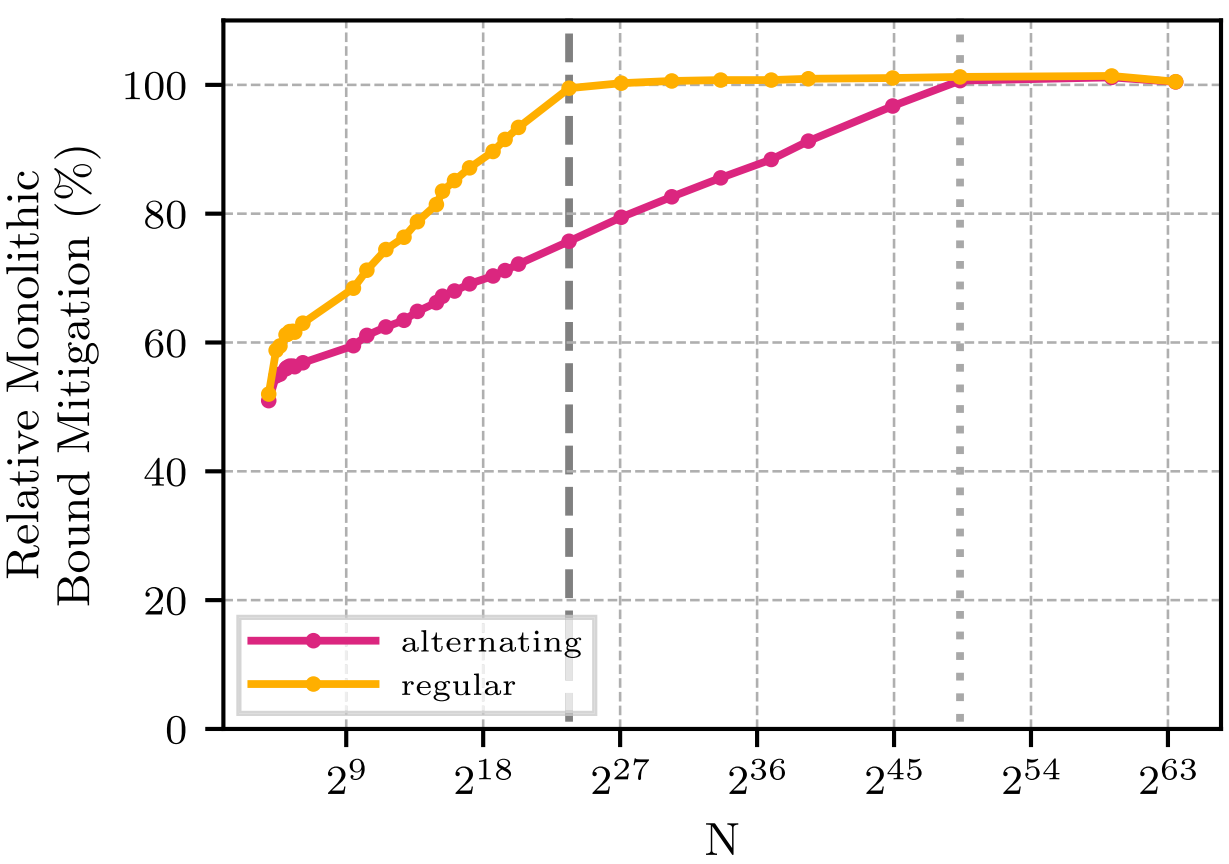

**FIGURE 23.** **Relative delay mitigation compared to the idle time bound of (18) for neutral atom QPU weighting.**

The double-iterative design performs slightly better, probably due to the described effect that the reordering of *CU* operations is leading to slightly looser phase processing dependencies. The three-cyclic approach, using three data qubits, even further improves over the iterative design for smaller circuits, as expected and performs close to the regular approach utilizing the full 3m qubits. In contrast to the standard variant, for the smallest circuits, the three-cyclic and regular approaches have an even better delay reduction, and the peak in improvement for the regular approach is reached earlier at $n = 12$. We argue that this effect stems from the two separate QFT of the data registers compared to the single QFT of the standard variant. As there are fewer dependencies for phase corrections based on measurement results, more time-consuming measurements can be executed in parallel.

To further analyze the delay reduction behavior, we break up the delay into two components: mitigable idle time and nonmitigable operational delay (*CU* executions and final phase processing). We isolated the mitigable portion by comparing the observed delay reduction to the theoretical bound for idle time, $t_{\text{idle}}$ from (18), as shown in Fig. 23. For small $N$, the mitigable phase processing delay dominates, but only a small fraction of idle time can be reduced via parallelization due to short *CU* operations. As $N$ increases, larger *CU* operations introduce more delay, enabling both alternating and regular designs to mitigate a greater portion of the idle time. The performance of the regular design peaks at $n = 25$, where it mitigates all idle time, while the alternating design achieves this at $n = 50$.

The trends in Fig. 22(a) reflect this interplay between mitigable idle time and nonmitigable operational delay. For the alternating design, the increase in idle time mitigation matches the increasing *CU* operation delay, keeping the relative reduction constant. In contrast, the regular design initially mitigates idle time faster than *CU* delay grows, leading to a greater relative delay reduction until it reaches the limit at $n = 25$. Beyond this, the diminishing significance of idle time compared to $CU$ delay causes a decrease in relative delay reduction.

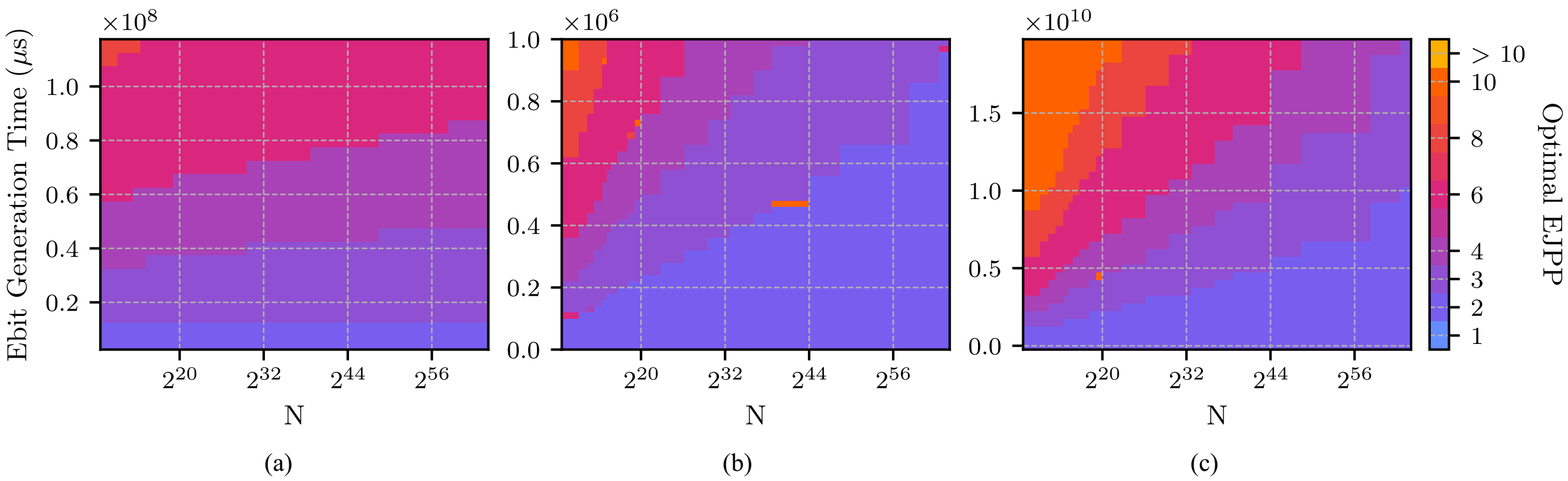

**FIGURE 24.** **Distributed delay scaling behavior for different ebit generation times and number *N* to factor. The heatmaps show the optimal choice for the number of ebit channels (denoted as EJPP) for the (a) neutral atom, (b) IBM superconducting, and (c) IonQ ion-trap weights.**

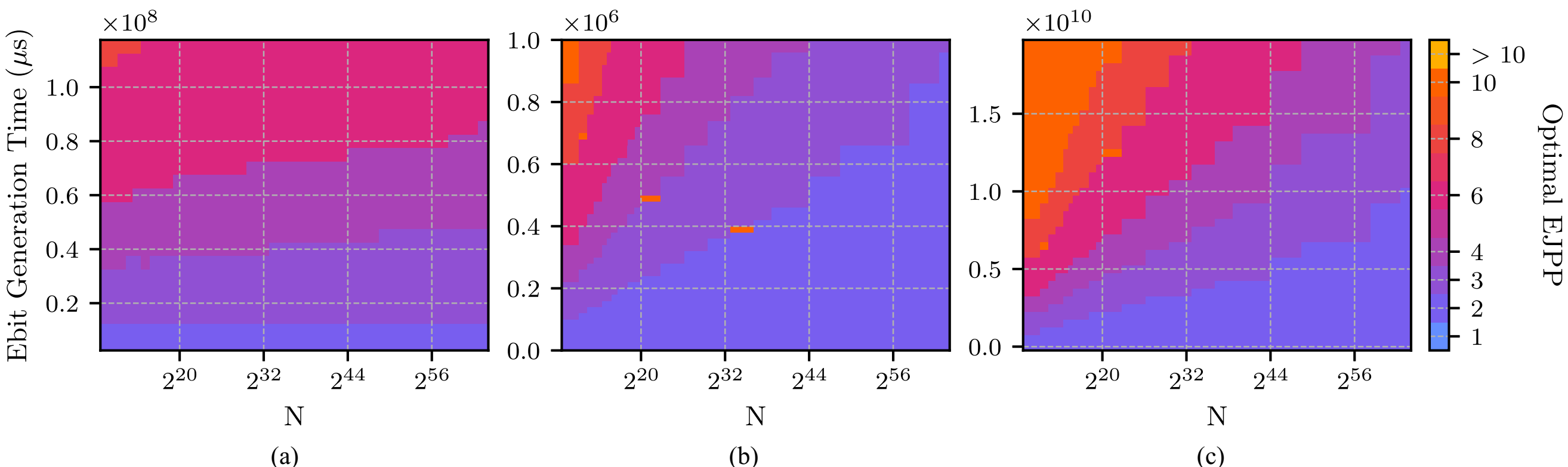

**FIGURE 25.** **Distributed delay scaling behavior for different ebit generation times and number *N* to factor. The heatmaps show the optimal choice for the number of ebit channels (denoted as EJPP) for the discrete logarithm variant of Shor's algorithm and the (a) neutral atom, (b) IBM superconducting, and (c) IonQ ion-trap weights.**

These results highlight the importance of tailoring design choices to the QPU characteristics and the size of $N$. For superconducting and ion-trap systems, the iterative design consistently performs best, as it is the most qubit-efficient approach and all designs have similar delays. For neutral atoms, the regular design is advantageous for smaller $N$, while for larger $N$, the proposed designs offer comparable delay efficiency to the regular design with the added benefit of requiring only one additional qubit over the iterative approach. When using the discrete logarithm variant, for small $N$, the three-cyclic design is preferred, while for larger $N$, the alternating or double-iterative designs are sufficient for optimal idle-time mitigation.

### B. ANALYSIS IN THE DISTRIBUTED SETUP

For the distributed parallelization, we extended the monolithic setup by integrating two QPUs connected via ebit channel counts from $\{1, 2, 3, 4, 6, 8, 10\}$. The circuits for all the selected values of $N$ from the monolithic experiments were recompiled, with their work and data registers partitioned across separate QPUs. We then evaluated the circuit delay across architectures where the ebit generation times were determined based on the hardware specifications outlined in Section IV-B.

Figs. 24 and 25 provide an overview of all conducted experiments. For a given value of $N$ and ebit generation time $t_{\text{ebit}}$, the delay of compiled circuits using one to four ebit channels is compared. The higher number of ebit channels is preferred only if it results in a lower circuit delay compared to configurations with fewer channels, even by a small margin. This approach reflects the high cost of ebit channels, emphasizing that their increased usage should be justified by more than a slight delay reduction.

We can also see from Fig. 24 that faster ebit generation times require fewer ebit channels, as the same ebit generation rate can be achieved with fewer channels, which also holds for the discrete logarithm variant, as shown in Fig. 25. In addition, for larger values of $N$, fewer ebit channels are needed because the required ebit generation rate decreases; slower ebit generation becomes acceptable since $CU$ operations take longer to complete.

In the ion-trap and superconducting setups, distributing the three monolithic design choices shows no substantial differences, which aligns with the monolithic analysis where all three approaches exhibited similar delays. Since in the monolithic setup the critical path is either only going through the $CU$ operations or the increase in delay when going through the phase processing operations is neglectable, new adaptations of the critical path due to distribution affect all designs equally. Therefore, in the following, we focus on the qubit-efficient alternating design for these two qubit technologies. The observed idle times and optimal choices of ebit channel counts are then closely approximated by our upper bounds of (24), as shown in Fig. 29(a) for the superconducting setup for the standard variant and the ion-trap setup for the discrete logarithm variant in Fig. 29(b).

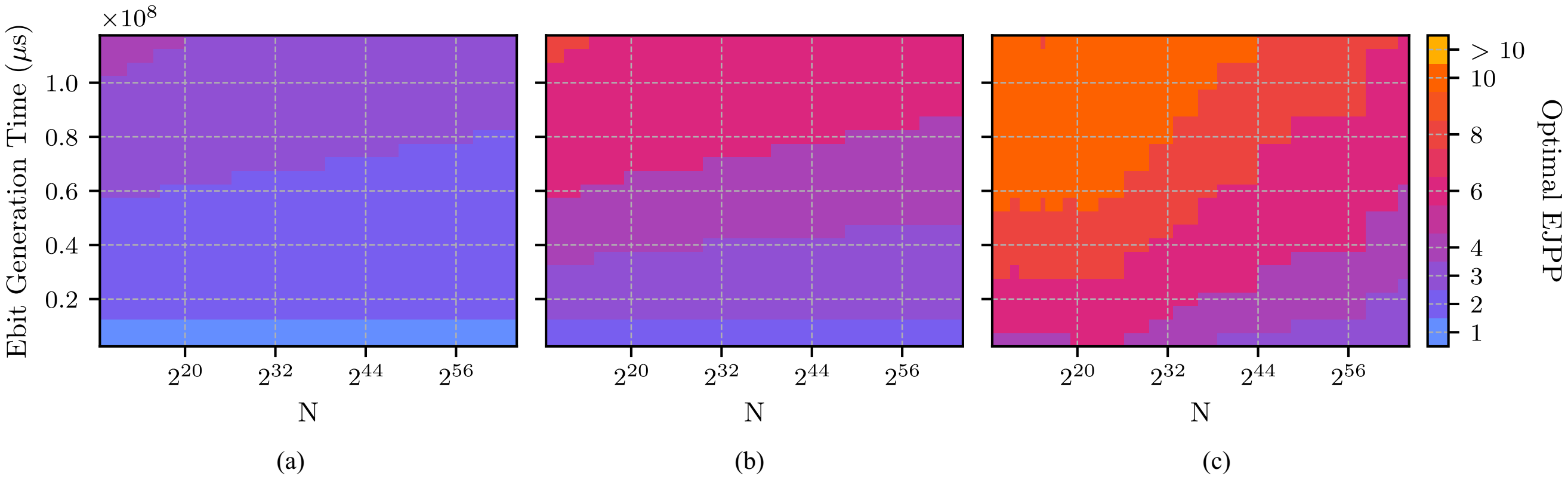

**FIGURE 26.** Distributed delay scaling behavior for different ebit generation times and number $N$ to factor. The heatmaps show the optimal choice for the number of ebit channels (denoted as EJPP) for the neutral atom setup and the (a) iterative, (b) alternating, and (c) regular design. The difference in phase processing idle time from the monolithic designs leads to different behaviors when distributed.

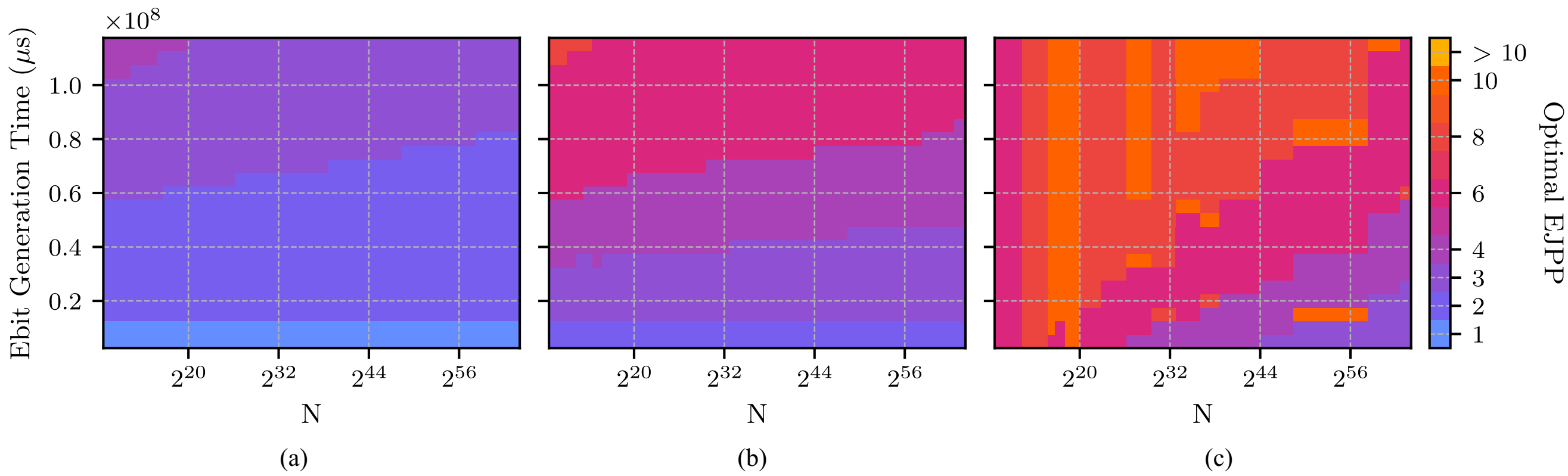

**FIGURE 27.** Distributed delay scaling behavior for different ebit generation times and number $N$ to factor using the discrete logarithm variant of Shor's algorithm. The heatmaps show the optimal choice for the number of ebit channels (denoted as EJPP) for the neutral atom setup and the (a) iterative, (b) alternating, and (c) regular design. The difference in phase processing idle time from the monolithic designs leads to different behaviors when distributed.

However, the neutral atom setup behaves differently, as illustrated in Figs. 26 and 27. In the iterative approach, fewer ebit channels are required compared to the alternating and regular cases. With very fast ebits, even a single ebit channel is sufficient. Conversely, the regular approach typically requires four ebit channels, with three being sufficient for larger $N$ and fast ebits. The alternating case falls in between, requiring two ebit channels for fast ebits and gradually more for slower ebits. This behavior is linked to the parallel execution of ebit generation and phase processing during $CU$ operations. In the iterative setup, phase processing takes a longer time, allowing ebit generation to be slower without causing idle time. In contrast, the regular setup minimizes idle time during phase processing, necessitating faster ebit generation to avoid distribution-related idle time. For the discrete logarithm-specific double-iterative design, the distribution behavior mostly replicates the behavior of the distributed alternating design [see Fig. 28(a)], though with slight differences and an outlier around $n = 15$ bits. We argue that these differences stem from the reordering of operations, as well as variances from the choice of $N$, $a$, and from the circuit transpilation and optimization behavior. Finally, the three-cyclic design requires more ebit channels than the alternating and double-iterative design [see Fig. 28(b)]. As the approach further reduces idle time due to phase processing, a higher rate of ebit generations is required. The more data qubits are used, the less idle time from phase processing is introduced, and with the regular setup, the behavior converges toward the approximate behavior of Fig. 29(c).

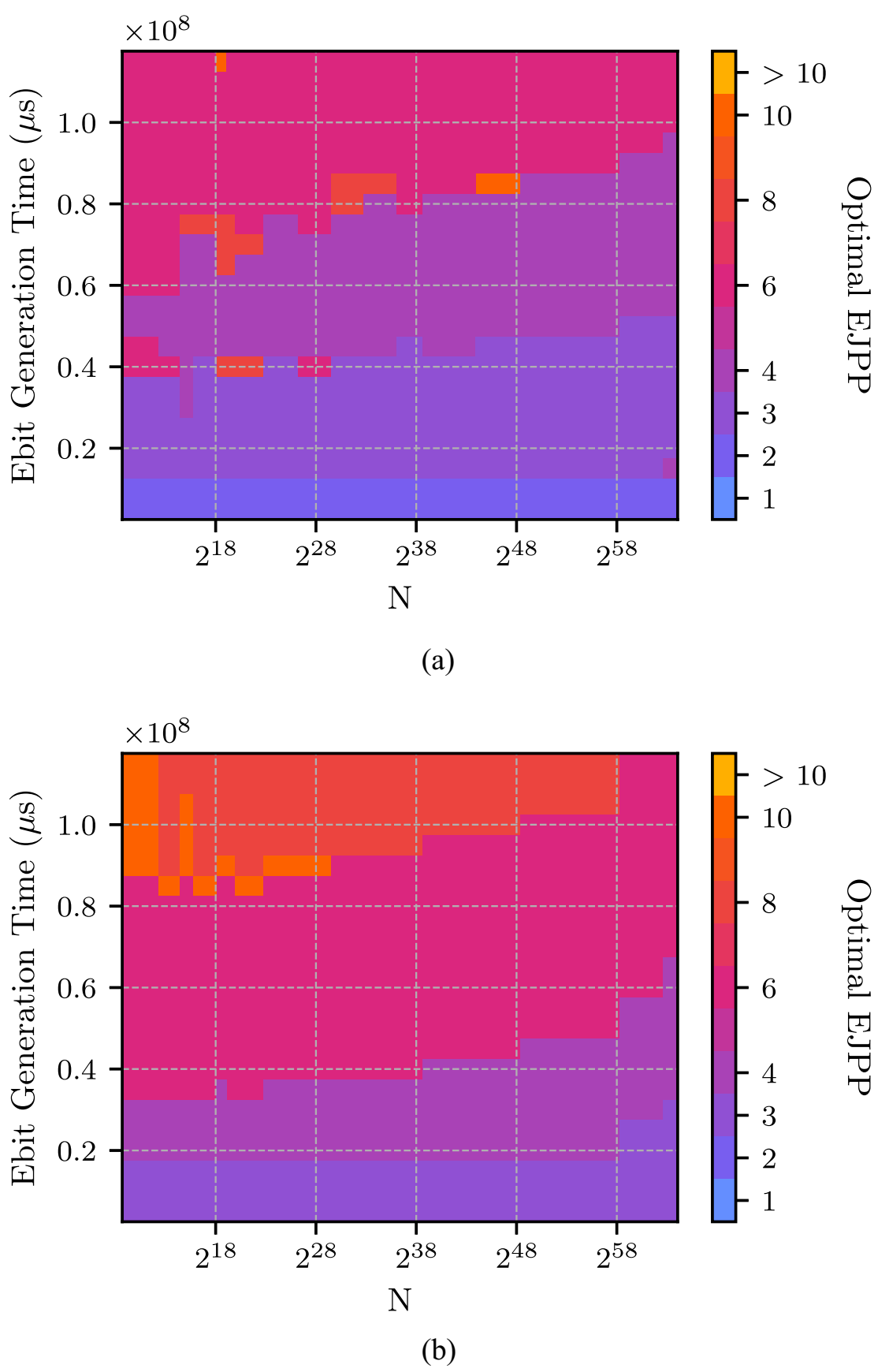

**FIGURE 28.** **Distributed delay scaling behavior for different ebit generation times and number *N* to factor using the discrete logarithm variant of Shor's algorithm. The heatmap shows the optimal choice for the number of ebit channels (denoted as EJPP) for the neutral atom setup using the (a) double-iterative and (b) three-cyclic design.**

From Fig. 29, we can see that with our configurations of up to ten ebit channels, we are reaching the maximum of ebit channels required to mitigate all distribution idle time for the superconducting QPUs, while for the ion-trap and neutral atom setups, more ebit channels would still be beneficial. Our experimental results from Figs. 25 and 27 show, though, that for regions in which larger ebit channel counts are required, in practice, smaller counts can already be sufficient and circuit delays are no longer majorly improved by higher ebit channel counts.

To analyze the transition points for optimal ebit channel configurations, we first examined the delay implications of multiple ebit channels by evaluating their behavior as $N$ scales, with fixed ebit times for a specific hardware setup. Fig. 30(a) illustrates the delay for one to four ebit channels in the alternating design with $t_{\text{ebit}} = 5.5$ s for the superconducting hardware configuration. Notably, the setup with a single ebit channel shows a substantially higher delay than the others. This is expected, as a single ebit channel offers no parallelization, resulting in prolonged idle times. Fig. 30(b) presents the reduction in delay relative to the single ebit channel setup. For small $N$, ebit generation constitutes the majority of the delay. With $k$ ebit channels, $k$ ebit generations can occur simultaneously, reducing the delay to approximately $1/k$ of that with no parallelization. However, as $N$ increases, the performance gains from additional ebit channels diminish.

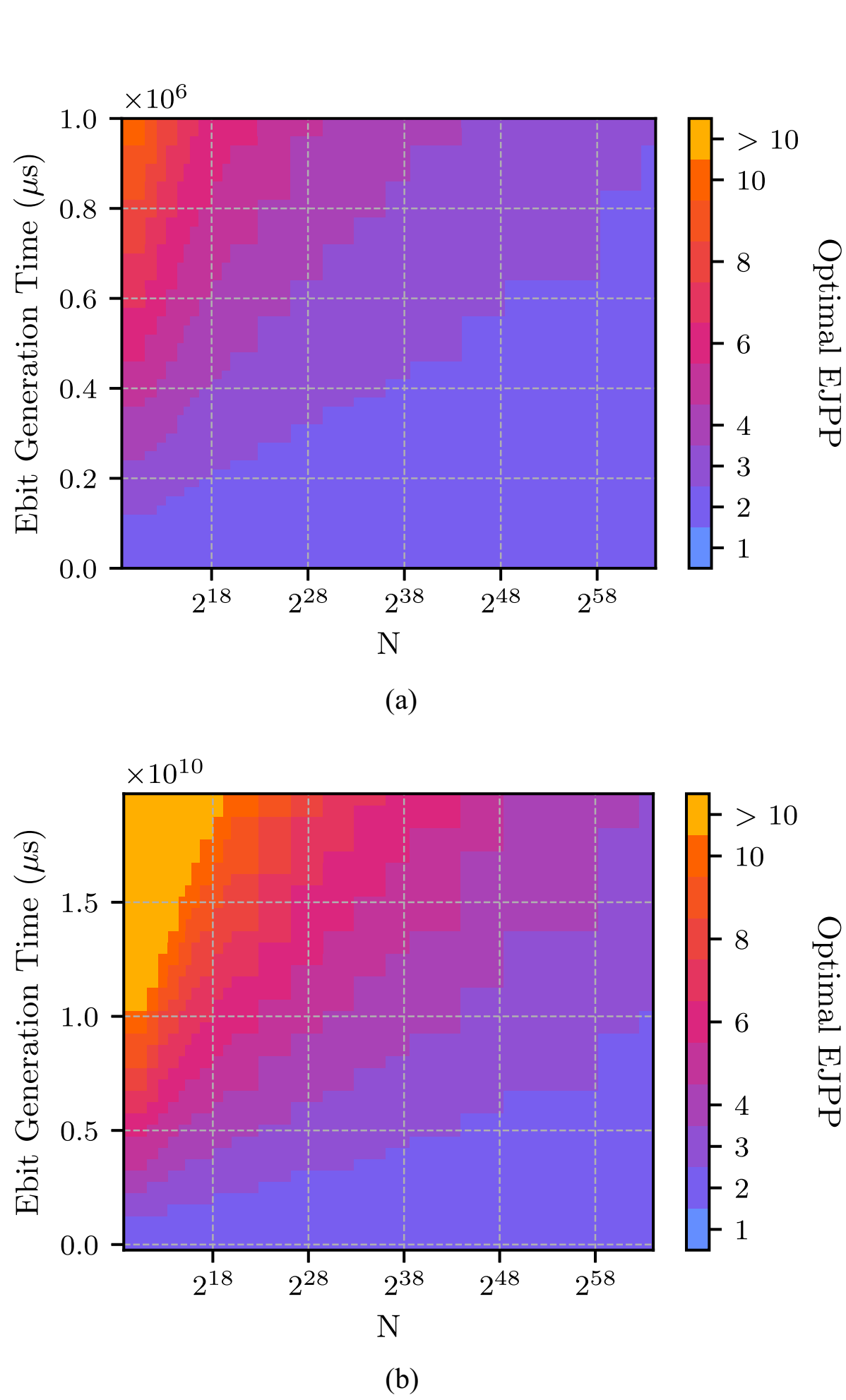

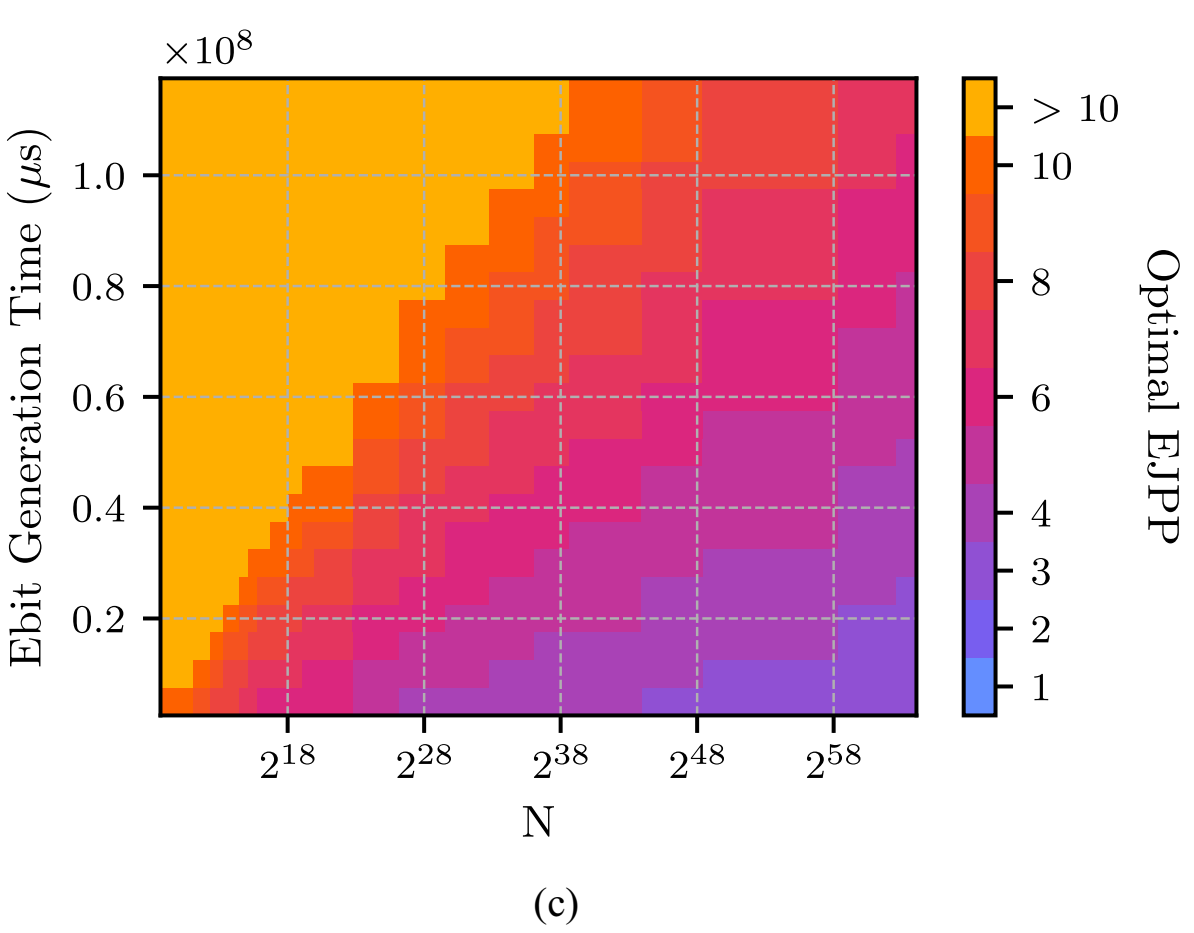

**FIGURE 29.** **Distributed delay scaling behavior for different ebit generation times and number *N* to factor. The heatmap shows the upper bound based on (24) for the number of ebit channels (denoted as EJPP) required to mitigate all idle time due to distribution for the superconducting setup of the (a) standard variant, as well as (b) the ion-trap and (c) neutral atom setup of the discrete logarithm variant.**

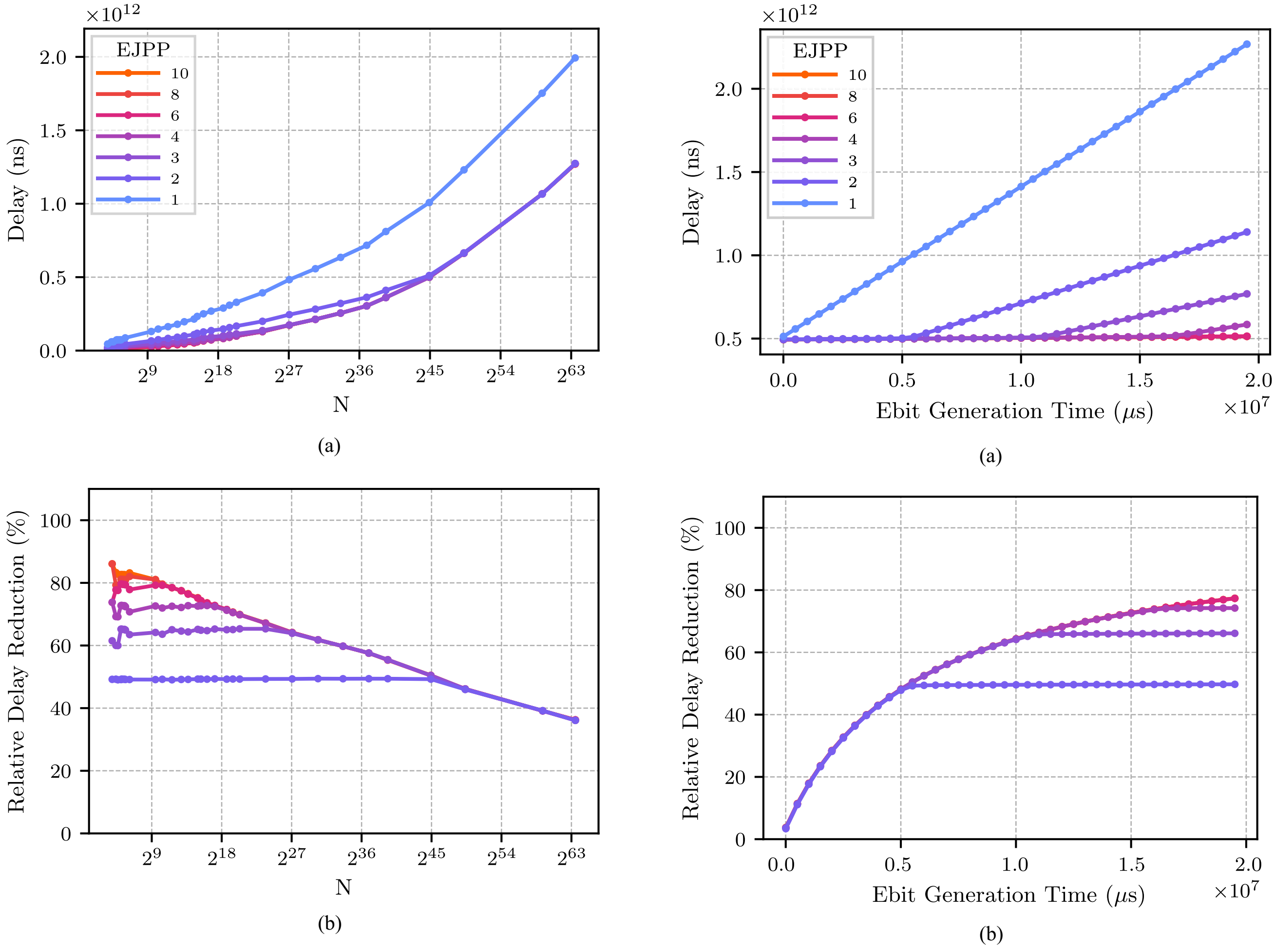

**FIGURE 30.** **Scaling of the (a) delay and (b) delay reduction for different *N* and fixed ebit generation time of 5.5 s for the ion-trap setup. (a) Scaling depth. (b) Scaling delay reduction.**

**FIGURE 31.** **Scaling of the (a) delay and (b) delay reduction for different ebit times and a fixed *N* with $n = 45$ bits for the ion-trap setup.**

We also explored a scenario where the hardware is not fixed, and instead, the number to be factored is predefined, requiring the selection of the optimal distribution hardware (ebit channel count and speed). Fig. 31 shows the delay scaling and relative delay reduction (compared to the delay with one ebit channel) for a fixed 45-bit number $N$. For very small ebit generation times, multiple ebit channels provide minimal benefit, as the impact of mitigable delay on the overall delay is negligible. However, as ebit generation times increase, parallelization becomes advantageous, with only two ebit channels being sufficient for effective parallelism. As seen in Fig. 31(b), when the ebit generation time approaches the average $t(CU)$ delay, three and four ebit channels start to outperform two. This is in line with the condition outlined in (23), where ebit generation exceeds the duration of a single $CU$ operation, making alternating between two channels inadequate. Similarly, when the ebit generation time approaches $2t(CU)$, three ebit channels no longer suffice, and four ebit channels are necessary to mitigate as much idle time as possible.

Fig. 32 demonstrates the extent to which parallelization mitigates the theoretical time bound [see (28)] for distributed computing, including ebit generation and the EJPP start- and end-processes. For small ebit generation times, setups with multiple ebit channels can eliminate all idle time except for the unavoidable initial ebit generation and start process. As the ebit generation time increases, we again observe the thresholds of $t(CU)$ and $2t(CU)$, where two and three ebit channels no longer provide optimal efficiency. For large ebit times, the time required for distribution dominates the delay and the results converge with those shown in Fig. 31(b). In this regime, with $k$ ebit channels, ebit generations are split into sets of $k$, which can be executed in parallel, reducing the circuit delay to $1/k$. This also reveals diminishing returns when moving from $k$ to $k + 1$ ebit channels.

These findings highlight how the performance of Shor's algorithm implementations is influenced by available hardware and problem size, providing valuable insights for making optimal design decisions. A midlevel perspective of multiple designs allows for more flexibility in leveraging targeted hardware efficiently. As our study shows, the difference in execution time between different designs and ebit channel parallelization strategies strongly depends on the hardware platform, problem size, and distribution setup. An analysis like ours can help determine how much of a targeted hardware's available compute resources can be effectively utilized for performance improvements. For instance, there may be a tradeoff between having multiple slower ebit channels or fewer faster channels, depending on the scale of Shor's algorithm being run. Other possible use cases are scenarios of multiple programs running on a single QPU or a QPU cluster. In such cases, a workload manager [68] could evaluate whether the benefit of more ebit channels for one program justifies limiting the resources available for others. Integrating our proposed analysis tools and designs into such tools would improve their allocation and scheduling capabilities.

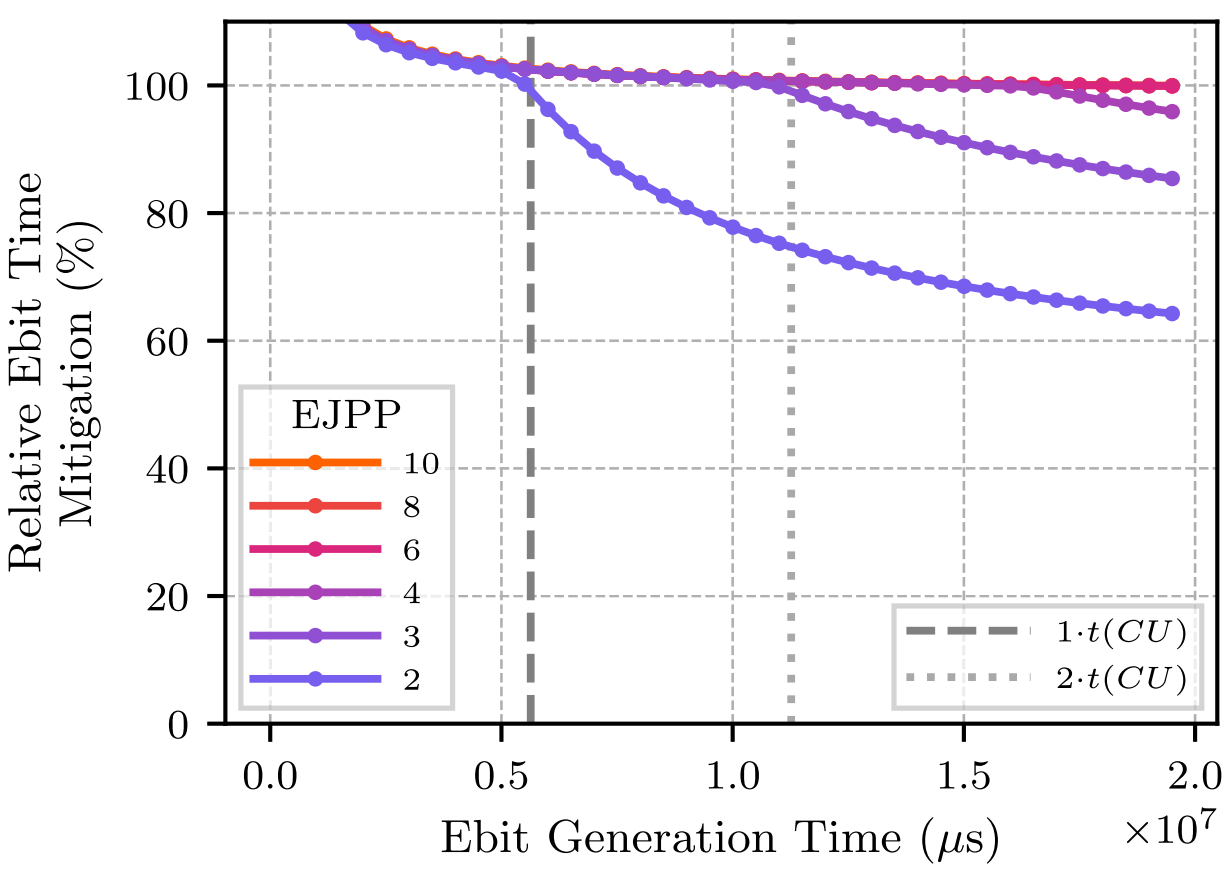

**FIGURE 32. Relative mitigation of idle time compared to the bound of idle time from distribution [see (24)] for different ebit times and a fixed *N* with *n* = 45 bits for the ion-trap setup. For small ebit times, circuit transpilation in overlaps between *CU* operations leads to mitigation above 100%.**

## VI. CONCLUSION

In this work, we investigated midlevel designs for Shor's algorithm from a timing perspective and introduced new designs for reducing idle time while preserving qubit efficiency. By reordering tasks to enable simultaneous execution, we improved overall execution time and extended these optimizations to distributed quantum systems. In distributed setups, we identified idle time bottlenecks caused by distribution protocols and proposed mitigation strategies using multiple ebit channels.

To systematically account for hardware-specific execution times, we adapted STA techniques from classical circuit design for quantum circuit optimization. This allowed us to incorporate gate execution delays and communication overhead in our design process. We evaluated our approaches on neutral atom, superconducting, and ion-trap platforms—both monolithic and distributed—by integrating modular arithmetic circuits from the QRISP framework. Our results confirmed that the proposed alternating, double-iterative, and three-cyclic designs effectively reduce idle time and demonstrated how hardware constraints can inform circuit optimization in monolithic and distributed settings.

Our approach complements existing arithmetic circuit optimization techniques by enhancing execution efficiency through task scheduling and idle time reduction. While prior research on Shor's algorithm has primarily focused on minimizing circuit depth and qubit count, we show that task reordering at a midlevel abstraction provides additional performance gains, especially when considering execution times of hardware.

Similarly, our midlevel approach to distributed execution does not replace necessary low-level task distribution but instead adds another layer of refinement for ordering tasks and managing information exchange. As large-scale quantum networks emerge, efficient scheduling and synchronization of distributed tasks will become increasingly critical.

Beyond Shor's algorithm, our findings suggest that similar parallelization and idle time reduction techniques could benefit other quantum algorithms with layered computational structures. QPE-based algorithms are a natural extension due to their structural similarities, but broader applications in quantum computing may also benefit from improved task scheduling and distribution-aware execution strategies.

Our study underscores the importance of hardware-aware quantum circuit design and serves as a foundation for integrating STA-inspired methods into quantum circuit compilation and design automation. A promising direction for future research is exploring additional timing-based optimization techniques for other quantum algorithms, particularly in error-corrected quantum systems, where logical operations introduce further time constraints.

An important next step is to validate our designs on real hardware once DQC systems become publicly available. While our study focuses on relatively large circuits, factoring 2048-bit RSA keys remains well beyond current capabilities. Reaching this scale will likely necessitate error-corrected quantum systems [5], [27], reinforcing the relevance of our techniques for logical quantum computing as a key area for future research.

Finally, as DQC advances, new challenges will arise in synchronization, task coordination, and ebit generation constraints. Our methods provide an initial step toward addressing these challenges by reducing idle time, and future research can further refine task scheduling and synchronization strategies to enhance distributed quantum execution, contributing to the broader goal of scalable distributed quantum computation.

## ACKNOWLEDGMENT

The authors thank Hans Hohenfeld, Gunnar Schönhoff, and Felix Wiebe for their helpful feedback. The authors also thank the anonymous reviewers for their constructive remarks and recommendations to improve this article.

*Author contributions:* Moritz Schmidt, Abhoy Kole, Leon Wichette, and Elie Mounzer conducted the research. Moritz Schmidt with the help from Leon Wichette wrote the code.

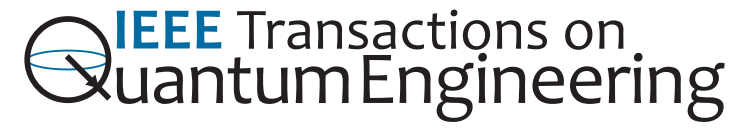

Abhoy Kole and Elie Mounzer formalized the approach. Elie Mounzer and Frank Kirchner supervised the work. Elie Mounzer provided directions. Moritz Schmidt drafted the initial manuscript and created the figures. All authors contributed to the manuscript's text and have read, reviewed, and approved the final manuscript.

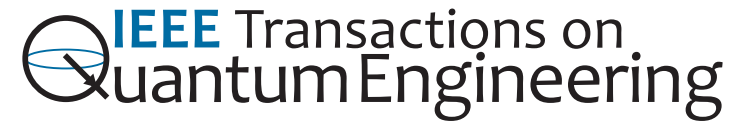

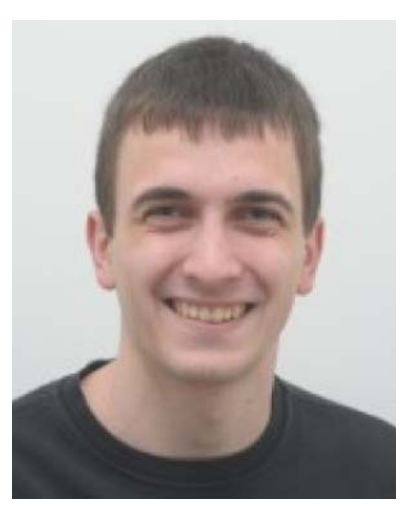

**Moritz Schmidt** received the B.S. degree in applied computer science from Ruhr University Bochum, Bochum, Germany, and the M.S. degree in computer science from the University of Stuttgart, Stuttgart, Germany, in 2018 and 2022, respectively. He is currently working toward the Ph.D. degree with the University of Bremen, Bremen, Germany.

His current research interests include distributed quantum computing and quantum error correction.

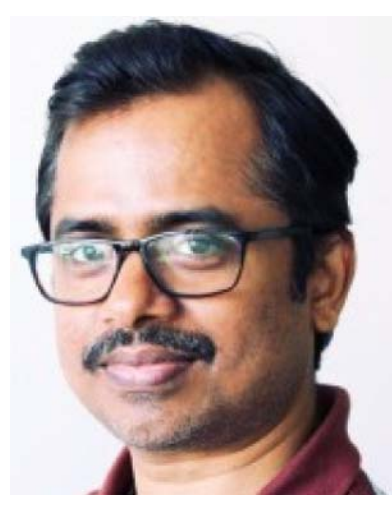

**Abhoy Kole** received the M.Tech. degree in information and communication technology from the Indian Institute of Technology Kharagpur, Kharagpur, India, in 2015, and the Ph.D. degree in computer science from the Rajendra Mishra School of Engineering Entrepreneurship, Indian Institute of Technology Kharagpur, in 2021.

He is currently a Postdoctoral Researcher with the Cyber-Physical Systems Group, German Research Center for Artificial Intelligence, Bremen. His current research interests include reversible and quantum computing, synthesis and optimization of quantum circuits, nearest neighbor realizations, and quantum fault tolerance.

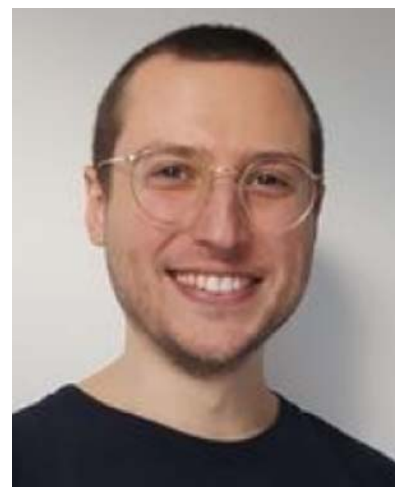

**Leon Wichette** received the B.S. and M.S. degrees in theoretical and mathematical physics from the University of Hamburg, Hamburg, Germany, in 2018 and 2022, respectively. He is currently working toward the Ph.D. degree with the German Research Center for Artificial Intelligence Robotics Innovation Center, Bremen, Germany.

His current research interests include distributed quantum computing and quantum error correction.

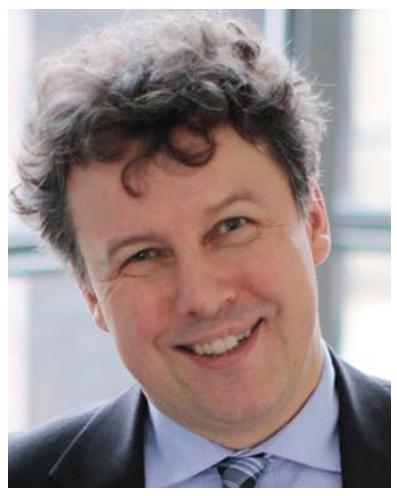

**Rolf Drechsler** (Fellow, IEEE) received the Diploma and Dr.phil.nat. degrees in computer science from the Johann Wolfgang Goethe University Frankfurt am Main, Frankfurt, Germany, in 1992 and 1995, respectively.

He was with the Institute of Computer Science, Albert-Ludwigs University, Freiburg im Breisgau, Germany, from 1995 to 2000, and with the Corporate Technology Department, Siemens AG, Munich, Germany, from 2000 to 2001. Since 2001, he has been a Full Professor and Head of the Group of Computer Architecture, Institute of Computer Science, University of Bremen, Bremen, Germany, where he was the Dean of the Faculty of Mathematics and Computer Science from 2018 to 2025. From 2008 to 2013, he was the Vice Rector for Research and Young Academics with the University of Bremen. In 2011, he also became the Director of the Cyber-Physical Systems Group, German Research Center for Artificial Intelligence, Bremen. His current research interests include the development and design of data structures and algorithms with a focus on circuit and system design.

Dr. Drechsler is a Fellow of the Association for Computing Machinery.

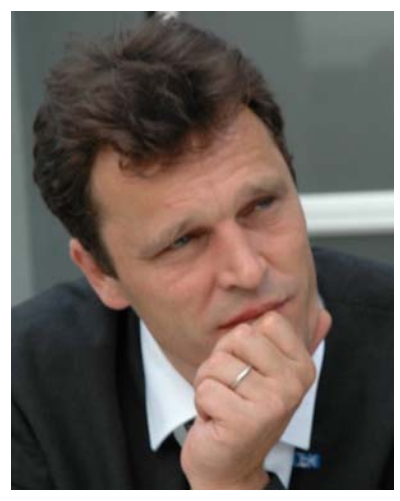

**Frank Kirchner** received the Diploma in computer science and neurobiology and the Dr.rer.nat. degree in computer science from the University of Bonn, Bonn, Germany, in 1994 and 1999, respectively.

Since 1994, he has been a Senior Scientist with the Gesellschaft für Mathematik und Datenverarbeitung, Sankt Augustin, Germany. In 1998, he joined, as a Senior Scientist, the Department for Electrical Engineering, Northeastern University, Boston, MA, USA, where he was appointed as an Adjunct and then a Tenure Track Assistant Professor in 1999. Since 2002, he has been a Full Professor with the University of Bremen, Bremen, Germany. Since December 2005, he has been the Director of the German Research Center for Artificial Intelligence Robotics Innovation Center, Bremen.

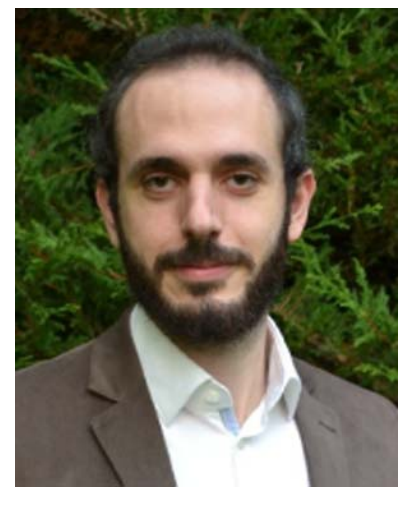

**Elie Mounzer** received the B.S. degree in physics from Lebanese University, Beirut, Lebanon, in 2015, and the M.S. and Ph.D. degrees in theoretical and mathematical physics from Paris-Saclay University, Orsay, France, in 2017 and 2022, respectively.

He is currently co-lead of the Quantum Computing Team with the German Research Center for Artificial Intelligence Robotics Innovation Center, Bremen, Germany. His current research interests include a variety of quantum computing topics, including quantum-inspired methods, quantum optimal control, distributed quantum computing, and quantum error correction.